\documentclass[conference]{IEEEtran}
\IEEEoverridecommandlockouts
\usepackage{cite}
\usepackage{placeins}
\usepackage{amsmath,amssymb,amsfonts}
\newtheorem{definition}{Definition}
\newtheorem{lemma}{Lemma}
\usepackage{listings}
\usepackage{amsmath,amsfonts,amssymb}
\usepackage{graphicx}
\usepackage{caption} 
\usepackage{afterpage}
\usepackage{longtable}
\usepackage{float}
\usepackage{verbatim}
\usepackage{stfloats}
\usepackage{longtable, booktabs, graphicx}
\usepackage{tabularx,booktabs}
\usepackage{lipsum}
\usepackage[noend]{algpseudocode}
\usepackage{ragged2e}
\usepackage{graphicx}
\usepackage[ruled,vlined,noline]{algorithm2e} 

\usepackage{cuted}
\usepackage{float}
\newtheorem{theorem}{Theorem}

\usepackage{textcomp}
\usepackage{xcolor}
\usepackage{subfigure}
\usepackage[english]{babel}
\usepackage[utf8]{inputenc}
\usepackage[nottoc]{tocbibind}
\def\BibTeX{{\rm B\kern-.05em{\sc i\kern-.025em b}\kern-.08em
T\kern-.1667em\lower.7ex\hbox{E}\kern-.125emX}}

\UseRawInputEncoding
\begin{document}
\title{Diffusion Models for Influence Maximization on Temporal Networks: A Guide to Make the Best Choice\\
{}
}
\author{
    \IEEEauthorblockN{Aaqib Zahoor, Iqra Altaf Gillani, Janibul Bashir} \\
    \IEEEauthorblockA{\textit{Department of IT}, NIT Srinagar} \\
    \IEEEauthorblockA{\{aaqib\_phaite003, iqraaltaf, janibbashir\}@nitsri.ac.in}
}

\maketitle

\begin{abstract}

The increasing prominence of temporal networks in online social platforms and dynamic communication systems has made influence maximization a critical research area. Various diffusion models have been proposed to capture the spread of information, yet selecting the most suitable model for a given scenario remains challenging. This article provides a structured guide to making the best choice among diffusion models for influence maximization on temporal networks. We categorize existing models based on their underlying mechanisms and assess their effectiveness in different network settings. We analyze seed selection strategies, highlighting how the inherent properties of influence spread enable the development of efficient algorithms that can find near-optimal sets of influential nodes. By comparing key advancements, challenges, and practical applications, we offer a comprehensive roadmap for researchers and practitioners to navigate the landscape of temporal influence maximization effectively.
\end{abstract}

\begin{IEEEkeywords}
Temporal Networks, Information Diffusion,Seed Selection,Influence Maximization,Online Social Networks,
\end{IEEEkeywords}

\section{Introduction}
The modern information ecosystem is characterized by an explosive rate of creation, dissemination, and absorption, largely orchestrated by online social platforms – think Instagram's visual narratives, TikTok's viral trends, and Reddit's collective intelligence \cite{bakshy2012role}. These digital architectures have fundamentally altered how information propagates, wielding significant influence over public opinion and global affairs. Intriguingly, the underlying mechanisms of diffusion are equally potent in physical domains. The COVID-19 pandemic starkly illustrated this, demonstrating how the architecture of physical contact networks dictates the spread of pathogens. Epidemiological efforts focused intensely on mapping and disrupting these networks to contain the virus, directly impacting its transmission dynamics and highlighting the critical role of diffusion science in safeguarding public health and societal stability. These compelling examples, spanning both virtual and physical realms, underscore the far-reaching consequences of information and influence diffusion on our interconnected world. However. unlike static networks, real-world interactions evolve over time, making influence propagation reliant on both structure and timing. Grasping these temporal dynamics is key to optimizing influence spread in dynamic ecosystem.

\begin{figure*}[ht]
    \centering
    \includegraphics[width=0.99\textwidth]{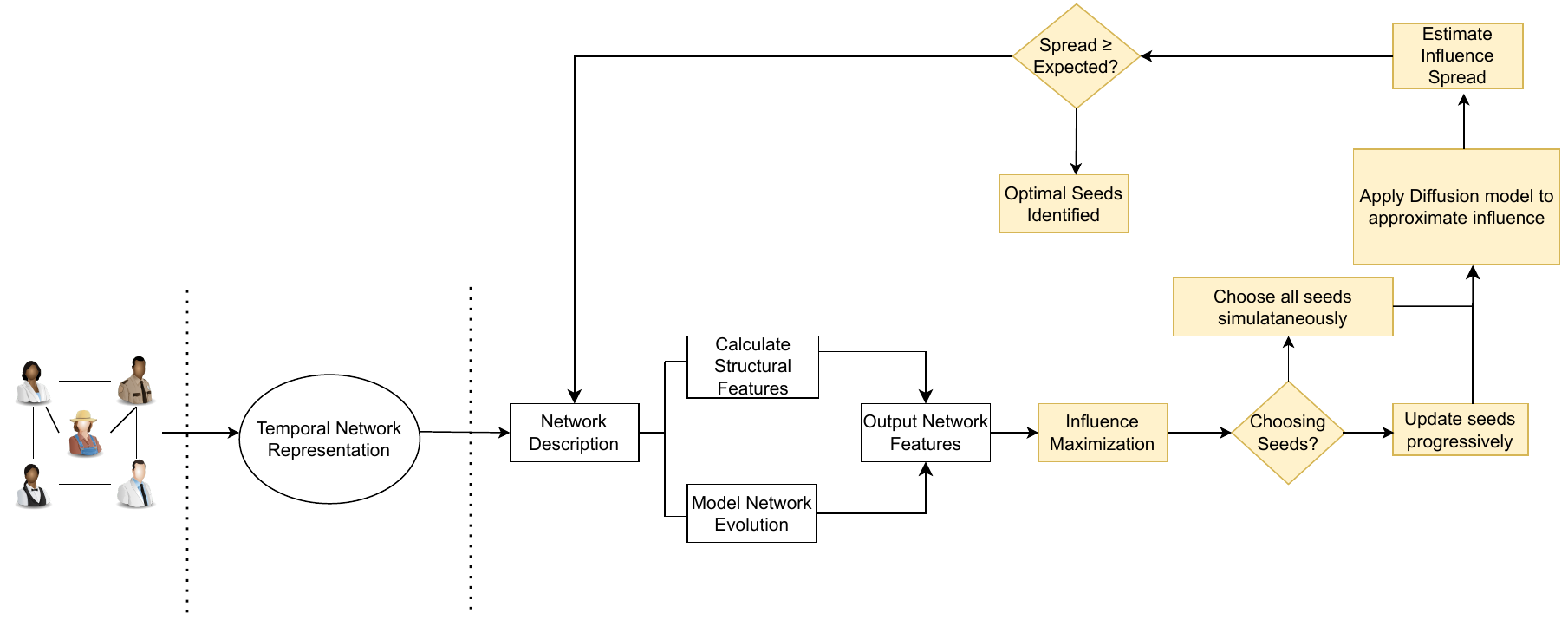}
    \caption{The influence research explores seed selection strategies and applies diffusion models to estimate the impact and spread of information.}
    \label{fig:mesh2}
\end{figure*}

Temporal networks, characterized by their dynamic and time-evolving connections, offer a more realistic framework for studying diffusion processes compared to static network models \cite{holme2012temporal}\cite{holme2015modern}. Unlike static networks, temporal networks capture the intricate ebb and flow of interactions, reflecting real-world complexities such as time-dependent activity levels, evolving relationships, and bursty interaction patterns. The limitations of static models become particularly apparent when considering scenarios where the timing and order of interactions are critical. For instance, in disaster relief efforts, the timely dissemination of emergency alerts through mobile communication networks can be a matter of life and death. The connections in these networks, calls, texts, data exchanges are inherently temporal, existing only during the specific moments of interaction and causing the network structure to constantly change \cite{zhang2016dynamics}. Similarly, the sustained momentum and global coordination of movements like BlackLivesMatter heavily rely on the precise timing of interactions on platforms like Twitter. These real-world examples underscore the necessity of moving beyond static representations to temporal networks, which can capture these crucial time-dependent dynamics. Representing these ever-changing interactions poses a significant challenge due to their diversity and complexity. Broadly, these representations are classified into two categories \cite{holme2012temporal}. In one class, to effectively incorporate the crucial timing information of interactions, some structural details of the network might be simplified or aggregated, leading to a lossy representation. This approach prioritizes the temporal ordering of events, potentially at the expense of fine-grained structural accuracy. In the other class, lossless representations aim to preserve all the details of both the network structure and the precise timing of each interaction. While offering a complete picture, these representations often come with increased complexity in terms of storage and computational analysis. Figure~\ref{fig:mesh2} presents the temporal network framework, which models time-resolved interactions using various representations \cite{holme2012temporal}. These representations facilitate a range of dynamic applications, including influence maximization, where seed nodes are selected either in a single phase or iteratively over time. The central focus of this study is on the influence maximization problem, as emphasized in the highlighted segment of the figure.
At the core of seed selection and influence maximization lie diffusion models, which govern how information, behaviors, and epidemics propagate through temporal networks.

Theoretical models of information diffusion, such as the Susceptible-Infected (SI), Independent Cascade (IC), and Linear Threshold (LT) models, provide valuable insights into how information propagates through networks based on interaction patterns \cite{pastor2015epidemic}. These models capture the mechanisms governing the spread of influence, offering a foundation for understanding large-scale diffusion processes. Despite their utility, several fundamental questions remain unanswered:

\begin{itemize}\item Which diffusion model best suits a particular application scenario?\item Which individuals or communities are most influential in driving large-scale information cascades?\item How do temporal interaction patterns affect the efficiency of information spread?\item What strategies can be employed to amplify or suppress specific diffusion processes?\item What are the effects on influence spread if the diffusion model is misspecified?\end{itemize}

Addressing these questions is essential for diverse applications, including enhancing public health outreach, mitigating the spread of harmful rumors, and optimizing marketing campaigns.

While considerable progress has been made in modeling diffusion, significant challenges remain, especially in the context of temporal networks. Real-world interactions evolve over time, and traditional models often overlook the importance of when connections occur. Recent research has begun to address this limitation by incorporating temporal motifs \cite{liu2021temporal}\cite{paranjape2017motifs}, which capture recurring patterns of interactions across time. These studies demonstrate that the timing, frequency, and alignment of interactions between individuals, rather than their sheer volume, play a crucial role in shaping the effectiveness of information spread. Recognizing these temporal dynamics is vital for developing realistic and effective diffusion strategies.

To build on these advancements, several surveys have systematically reviewed diffusion models, offering insights into their theoretical foundations, applications, and computational strategies while highlighting key challenges in temporal network modeling.
\subsection{Existing Surveys}
The study of information diffusion modeling can be divided into two primary methodologies: time-series and data-driven approaches. Time-series models emphasize mathematical formulations derived from diffusion data, providing clear interpretations and predictions of information spread over time. These models encompass differential and difference equations for volume predictions, individual adoption prediction frameworks (both progressive and non-progressive), and likelihood maximization techniques for understanding propagation relationships. Representing the classical approach to diffusion analysis, these models are well-defined and interpretable. In contrast, data-driven models leverage machine learning (ML) algorithms to learn patterns directly from data, capturing the underlying dynamics of diffusion without explicit modeling. Advances in data availability and computational power, coupled with the integration of technologies such as Natural Language Processing (NLP), have significantly enhanced these models' capabilities, enabling the direct learning of complex content semantics and redefining traditional boundaries of model design.

While prior surveys \cite{jiang2016identifying}\cite{li2017survey}\cite{wang2013modeling}\cite{gao2019taxonomy}\cite{guille2013information}\cite{li2018influence}\cite{singh2019survey}\cite{yao2015diffusion}\cite{yuan2016will}\cite{he2016cost}\cite{wang2012diffusive}\cite{10.1145/3485273} have explored various aspects of information diffusion models, this review distinguishes itself by addressing fundamental questions as discussed before and providing actionable insights. Many surveys focus narrowly on specific applications such as source detection \cite{jiang2016identifying}, worm propagation \cite{wang2013modeling}, or influence maximization \cite{li2018influence}, often neglecting a thorough overview of foundational diffusion models. Additionally, certain reviews exclude critical aspects such as relationship inference and non-progressive models \cite{gao2019taxonomy}\cite{guille2013information}\cite{li2017survey}\cite{singh2019survey}, which are integral to understanding comprehensive diffusion processes. The classification criteria traditionally employed, whether predictive or explanatory, or based on topological factors, often fall short in describing the nuances of emerging models that blend these boundaries, such as models embedding data into a unified space to predict diffusion paths based on geometric relations among information and users. Moreover, many existing models treat the information diffusion process in isolation, disregarding the interactivity and influence of social behavior \cite{he2016cost}\cite{wang2012diffusive} and excluding the comprehensive measurement of diffusion models \cite{10.1145/3485273}, despite the clear impact social vectors and user interactions have on information spread.

Our survey bridges these gaps by offering a comprehensive guide to selecting the most suitable diffusion model for specific application needs, creating a taxonomy that categorizes various diffusion models on static and temporal networks. By integrating these elements, our review not only synthesizes the current landscape of diffusion models but also provides actionable insights into model selection, evaluation, and optimization, setting it apart from existing literature.

\subsection{Our Contributions}  

This article advances the study of information diffusion on temporal networks through the following key contributions:  

\begin{itemize}  
    \item \textbf{Comprehensive Taxonomy of Diffusion Models:} We propose a structured classification of existing diffusion models, categorizing them based on their theoretical foundations, temporal dynamics, and applicability to real-world scenarios. This taxonomy provides a systematic understanding of model characteristics, facilitating informed decision-making for researchers and practitioners.  

    \item \textbf{Systematic Framework for Influence Maximization:} We introduce a structured methodology for selecting diffusion models based on two critical objectives: maximizing influence spread and minimizing computational overhead. Our framework categorizes models according to network properties, adaptive thresholds, competitive dynamics, and higher-order interactions, while also identifying scalable solutions with provable guarantees, real-time optimizations, and hybrid approaches that balance efficiency and effectiveness.  

    \item \textbf{Principled Model Selection Guidance:} Building on our framework, we present an in-depth guide to help researchers and practitioners choose diffusion models best suited to specific application contexts. To enhance practical understanding, we provide concrete use cases that demonstrate how different models perform under various real-world conditions. By aligning selection criteria with influence maximization objectives, we bridge the gap between theoretical advancements and practical deployment, ensuring optimal model performance in diverse temporal network settings.  
\end{itemize}

\subsection{Organization}

This article is designed to provide a structured methodology for selecting the most appropriate diffusion model based on the findings presented in this research. To facilitate this process, we provide a detailed flowchart (see Figure 2), which serves as a decision-support tool for researchers and practitioners. The flowchart begins with a formal problem definition, distinguishing between general influence maximization problems and specific target-based applications. It then guides users to classify their problem domain by referencing the taxonomy of diffusion models outlined in Section V. Depending on the prioritization of objectives, whether minimizing computational overhead or maximizing diffusion spread, the flowchart directs users to consult either Section VI A (for computational efficiency optimization) or Section VI B (for spread maximization strategies). We provide application scenarios and use cases in Sections VII and VIII, where any application scenario can be mapped to the given cases to identify and select the optimal diffusion model for a given application context.

\begin{figure*}[ht]
    \centering
    \includegraphics[width=\linewidth]{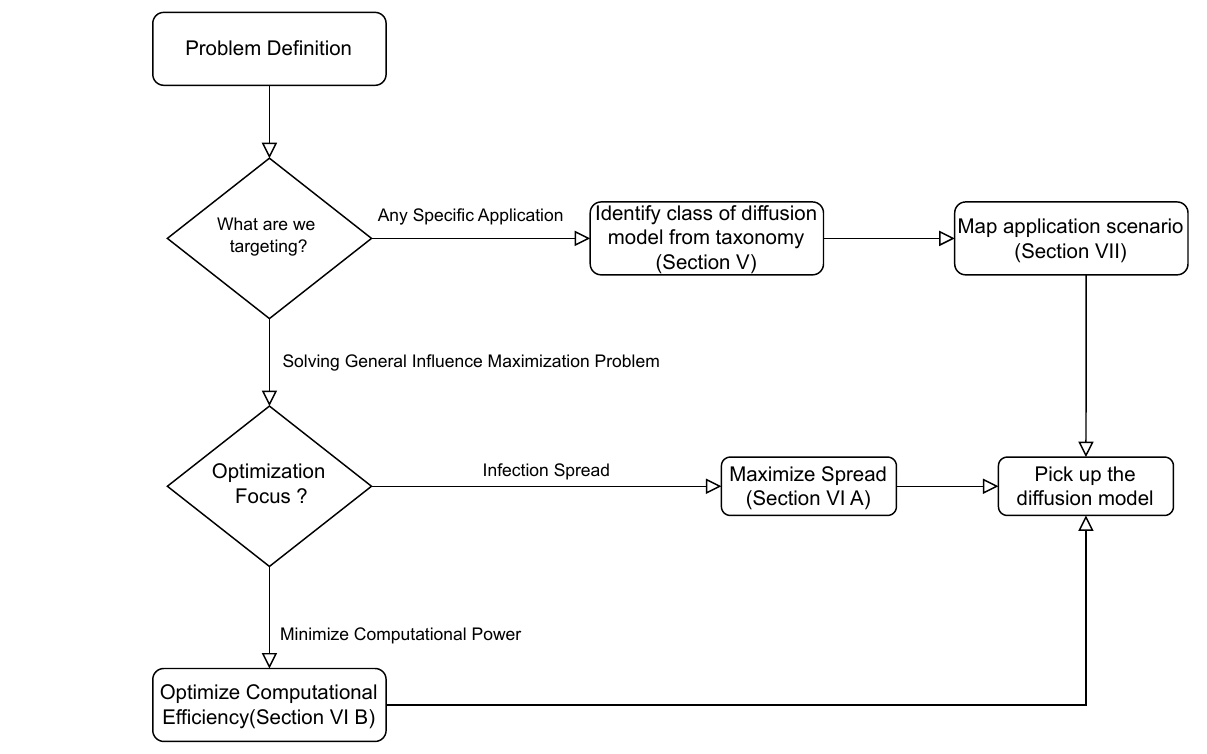}  
    \caption{Flowchart showing how to pick up right diffusion model for given requirements}
    \label{fig:mesh3}
\end{figure*}

\section{Preliminaries}  

A \textit{temporal network} is a dynamic graph where interactions between nodes evolve over time, capturing the temporal nature of relationships. Formally, it is represented as \( \mathcal{G}(V, E, T) \), where \( V \) is the set of nodes, \( E \subseteq V \times V \times T \) is the set of time-stamped edges, and \( T \) denotes the discrete or continuous timeline over which interactions occur. Each temporal edge \( e = (u, v, t) \) signifies an interaction from node \( u \) to node \( v \) at time \( t \), establishing a causal structure that constrains the flow of information. These networks serve as foundational models for diverse real-world processes, such as information diffusion, communication dynamics, biological interactions, and epidemic spreading \cite{holme2015modern, holme2012temporal}.  

The process of diffusion governs how information, influence, or contagion propagates through a temporal network. Formally, the influence spread can be quantified by a function \( \sigma: 2^V \to \mathbb{R}_{\geq 0} \), which measures the expected number of influenced nodes given an initial seed set \( S \). Unlike static networks, where influence spreads over fixed edges, temporal networks impose additional constraints; reachability depends not only on structural connectivity but also on the timing of interactions. The temporal sequence of edges plays a crucial role in determining the speed, extent, and efficiency of diffusion, making it essential to consider time-aware strategies when optimizing influence maximization.

\begin{definition}[Monotonicity]\label{prop:monotone}  
A function \( \sigma(S) \) is monotone if adding nodes to the seed set does not decrease the spread, i.e., for any \( S \subseteq T \subseteq V \):  
\begin{equation}  
    \sigma(S) \leq \sigma(T).  
\end{equation}  
\end{definition}  

\begin{definition}[Submodularity]\label{prop:submodular}  
A function \( \sigma(S) \) is submodular if the marginal gain from adding a node \( u \) to a smaller set \( S \) is at least as large as adding it to a larger set \( T \), i.e., for all \( S \subseteq T \subseteq V \) and \( u \in V \setminus T \):  
\begin{equation}  
    \sigma(S \cup \{u\}) - \sigma(S) \geq \sigma(T \cup \{u\}) - \sigma(T).  
\end{equation}  
\end{definition}  

\begin{definition}[Social Vectors]
Social vectors highlight the heterogeneity in social connections. Users may have multiple accounts, and physical-world friends might not be connected on Social Networking Service (SNS) based platforms. Different SNS platforms also offer various modes of interaction, such as public browsing on Twitter versus private messaging on WhatsApp.
\end{definition}

\begin{table}[htb!]
\centering
\begin{tabular}{ |p{2cm}|p{6cm}| }
 \hline
Symbol & Definition \\
 \hline
 $G(V, E, T)$ & Social network topology with user set \( V \), edge set \( E \), and time stamp \( T \) \\
 $C$ & Cascade set \\
 $c$ & A single cascade \\
 $\sigma_D(S, G)$ & Influence spread initiated by the seed set \( S \) based on a diffusion model \( D \) \\
 $w_c(t)$ & Weight of cascade \( c \) at time \( t \) \\
 $T$ & Temporal network \\
 $V_k^{t_k}$ & Node \( k \) at time stamp \( t_k \) \\
 $t_c^k$ & Infection time of the \( k \)-th forwarding in cascade \( c \) \\
 $u_c^k$ & \( k \)-th forwarding user in cascade \( c \) \\
 $\delta_{ij}$ & Kronecker delta function \\
 $N$ & Total number of nodes in the network \\
 $\rho_1, \rho_2$ & Fraction of red agents in respective cliques \\
 $\alpha$ & Proportion parameter in clique partitioning \\
 $\phi$ & Graph conductance, affecting consensus time \\
 $\psi$ & Reward associated with correct decision-making \\
 $r$ & Discount rate in expected utility calculation \\
 \( F(t) \) & Fraction of adopters (or aware individuals) at time \( t \) in the Bass model \\
 \( \pi_i(t) \) & Surplus (utility) of agent \( i \) at time \( t \), influenced by local network effects \\
 $I_t$ & Information available at time \( t \) \\
 $N(t)$ & Total user population at time \( t \) \\
 $S(t)$ & Susceptible (inactive) user set at time \( t \) \\
 $I(t)$ & Infected (active) user set at time \( t \) \\
 $\Delta I(t)$ & Newly infected user set at time \( t \) \\
 $D_{in}^u(t)$ & Incoming neighbor set of user \( u \) at time \( t \) \\
 $D_{out}^u(t)$ & Outgoing neighbor set of user \( u \) at time \( t \) \\
 $p_c^{uv}(t)$ & Propagation probability of cascade \( c \) from \( u \) to \( v \) at time \( t \) \\
 $A(t)$ & Transmission rate matrix at time \( t \) \\
 $f_u(m)$ & Feature vector of user \( u \) for topic/message \( m \) \\
 \hline
\end{tabular}
\caption{Basic notations}
\label{table:1}
\end{table}

\section{Influence Maximization}
Influence maximization is a fundamental problem in social network analysis, aimed at identifying a small subset of influential nodes (users) in a network to maximize the spread of influence. This problem has applications in viral marketing, rumor spreading, public health campaigns, and political mobilization. The core idea is that information, behaviors, or innovations propagate through networks in a cascading manner, much like the spread of a virus. By strategically selecting the initial set of influential users, organizations, and researchers can optimize the reach of their campaigns with minimal resources.

A crucial aspect of influence  maximization is modelling how influence propagates through a network. This is where diffusion models come into play. Diffusion models mathematically describe how information spreads across the network, determining how one node's activation leads to others being influenced. These models help formalize the process of information diffusion, allowing researchers to estimate the influence of a given set of nodes and develop efficient algorithms for influence maximization. Formally, the problem is defined as:
\begin{definition}[Influence Maximization]\label{def:IM}
Given a social network represented as a graph $G = (V, E)$, where $V$ is the set of nodes (users) and $E$ is the set of edges (social connections), the 	\textit{Influence Maximization (IM) Problem} seeks to identify a subset $S \subseteq V$ of size $k$ ($k < |V|$) such that the expected number of influenced users in the network, denoted by an influence function $\sigma_D(S, G)$, is maximized:
\begin{equation}
    \text{IMM}(G, k) = \arg\max_{S \subseteq V, |S| = k} \sigma_D(S, G)
\end{equation}
where $\sigma_D(S, G)$ models the spread of influence initiated by the seed set $S$ based on a diffusion model (e.g., Independent Cascade or Linear Threshold Model).
\end{definition}

In case of temporal networks, the selection of seeds is time dependent. Apart from the seed set S, we need to choose the time stamp $t_i$ at which each node $V_i \in S$ is activated so that it can influence the neighbors as per the diffusion model $D$ in time $t > t_i$. The nodes activated at time stamps t are added to the set of active nodes that further influence the neighborhood as per $D$. we can redefine the problem as: 
\begin{definition}[Influence maximization on temporal networks]
Given a temporal network $T$ and diffusion model $D$, find a set $S$ of nodes where $S={V_1^{t_1},V_2^{t_2},...V_k^{t_k} }$;$V_i \neq V_j$ to be activated such that $\sigma_D(S)$ is maximized.
\end{definition}

In the static setting, the simplest appraoch to find S that maximizes $\sigma_D(S)$ is by using the greedy algorithm that tries to choose nodes leading to optimal guarantee of overall gains in the influence spread. The approach yields a solution that is at maximum $(1 - \frac{1}{e})$ times far away from the global optimum \cite{kempe2003maximizing}. This is possible because the objective function is submodular and monotone in static network setting. The natural extension of greedy algorithm to temporal setting does not always guarantee optimal gap and is dependent on the diffusion model employed. The majority of the diffusion models on temporal networks render the objective function to be non-monotone and non-submodular, thereby posing the question of getting optimal guarantees through greedy optimization. We will discuss it further in seed selection mechanisms.

Even though by manipulating the diffusion models on temporal networks or using the SI model paves a way to extend greedy algorithms on temporal networks, the approach is computationally very expensive; therefore, resorting to heuristic methods saves us from hectic computations. But the problem with heuristic methods lies in the selection of the heuristic itself. It is difficult to choose a particular node ranking heuristic in a dynamic environment because we never know which ranking method is important in what scenario. Although a mixed heuristic methodology can work in such situations, for example choosing multiple heuristics like degree, centrality, similarity and calculating an overall score, there is still a problem of overlapping. That is, two nodes with similar neighbors can spread the influence to the same nodes, which results in wastage and is not an optimal solution. Therefore, to estimate the $\sigma_D(S)$ and select the seeds, there is a tradeoff between reducing computation time and ensuring the quality of selected seeds.
To better understand and manage this tradeoff, it is important to first examine the computational complexity associated with influence maximization. In the next section, we discuss a variety of seed selection mechanisms that achieve balance in this tradeoff.

\subsection{Computational Complexity and Algorithmic Approaches}

Computing the influence spread of a seed set is computationally intractable in general. Under both the Independent cascade(IC) and Linear Threshold(LT) models, estimating $\sigma(S)$ is \#P-hard. Moreover, the influence maximization problem itself has been proven NP-hard under IC and LT models. These hardness results imply that exact solutions are infeasible for large networks, necessitating the development of approximate or heuristic approaches. Fortunately, under classical models , the influence function $\sigma(S)$ is monotone and submodular, enabling a greedy algorithm that provides a $(1 - 1/e)$ approximation to the optimal solution. However, its reliance on costly Monte Carlo simulations for estimating marginal gains poses scalability challenges.

In temporal networks, the structure of the influence function is no longer guaranteed to be monotone or submodular, depending on how influence decays or propagates over time. This makes greedy algorithms less reliable and affects their theoretical guarantees. As a result, heuristic strategies are more commonly applied. Simple heuristics rank nodes based on structural metrics like degree, betweenness, or closeness, but these metrics may not capture the temporal intricacies of real-world diffusion. Mixed heuristics that combine several measures can offer improved performance but often suffer from influence overlap, where highly ranked nodes affect the same neighbors, leading to redundant selection.

Beyond heuristics, algorithmic frameworks like Reverse Influence Sampling (RIS) have shown remarkable performance in static networks by transforming influence estimation into a sampling problem. RIS-based algorithms significantly reduce runtime while retaining the approximation guarantees. Additionally, meta-heuristic methods such as genetic algorithms, particle swarm optimization, and simulated annealing have been explored for influence maximization, especially when constraints such as budgets, dynamic topologies, or time windows are involved. These algorithms provide flexible optimization capabilities, albeit at the cost of interpretability and reproducibility.  Importantly, the effectiveness of any influence maximization technique is deeply tied to the underlying diffusion behavior of information in the network, which we now explore in the following subsection.

\subsection{Information Diffusion}
Influence propagation describes the process through which information disseminates across a network as nodes share content with their neighbors, who, in turn, propagate it further. This phenomenon is particularly evident in viral marketing, where information spreads rapidly through social interactions, highlighting the significance of influence propagation in social networks. When individuals share content within their immediate connections, who subsequently forward it to others, the information reaches a broader audience in a cascading manner.

To model and predict how information and behaviors diffuse within these networks, diffusion models are essential. These models capture real-world propagation dynamics by considering various factors such as transmission probabilities, network topology, and temporal characteristics. Standard diffusion models are typically formulated based on observed patterns of information flow in real-world networks. They are then leveraged to simulate diffusion processes, aiding in the resolution of specific problems such as influence maximization.

For instance, during the COVID-19 pandemic, epidemiological models like the Susceptible-Infected-Recovered (SIR) model were employed to simulate virus transmission and assess the impact of interventions such as social distancing, mask mandates, and vaccination programs. These models played a crucial role in guiding policymakers by enabling timely interventions, thereby mitigating the spread of the virus and reducing the strain on healthcare systems.

It is important to recognize that diffusion processes are inherently time-dependent, as propagation occurs across nodes at different time steps. For example, in the Independent Cascade (IC) model, a node \( A \) may influence its neighbor \( B \) at time \( t_1 \), and \( B \) may subsequently influence node \( C \) at time \( t_2 \). However, this conceptual time in diffusion models differs from the temporal structure of real-world networks, where the existence of links between nodes varies over time. In such networks, the success of an influence attempt at a given time step is contingent upon the dynamic nature of network connectivity. This distinction has significant implications for influence maximization strategies. The optimal seed set for maximizing influence in a static network may differ substantially when the temporal nature of the network is considered.

Below, we provide a simple example to illustrate this concept. Consider a static network \( G = (V, E) \) with nodes \( V \) and edges \( E \). Let \( D \) be a node with high degree centrality, implying a large number of direct connections. Under the Independent Cascade (IC) model, the probability that \( D \) directly influences its neighbors is given by:

\[
P_{\text{static}}(D) = 1 - \prod_{j \in N_D} (1 - \beta),
\]

where \( N_D \) is the set of neighbors of \( D \), and \( \beta \) represents the transmission probability, the probability that an active node successfully influences an inactive neighbor in a single interaction. Typically, \( \beta \) depends on factors such as the type of interaction (e.g., social influence, epidemic spreading), edge weight, and the strength of the connection between nodes.

If \( D \) has three neighbors (\( B, C, E \)), the probability of influencing a single neighbor in one step is:

\[
P_{\text{static}}(D \rightarrow E) = \beta.
\]

Now consider a temporal network \( G(t) = (V, E(t)) \), where edges \( E(t) \) evolve over discrete time steps \( t_1, t_2, t_3, \ldots \). The probability that \( D \) influences its neighbors in this network depends on the sequence and timing of interactions. Assume that \( D \) can influence node \( E \) only through intermediate nodes \( B \) and \( C \) at times \( t_1 \) and \( t_2 \), respectively. The compounded probability under the temporal diffusion model is given by:

\[
P_{\text{temporal}}(D \rightarrow E) = \beta \cdot \beta \cdot \beta = \beta^3.
\]

For instance, if \( \beta = 0.5 \), then:

\[
P_{\text{static}}(D \rightarrow E) = 0.5,
\]

whereas in the temporal network,

\[
P_{\text{temporal}}(D \rightarrow E) = 0.5^3 = 0.125.
\]

This significant reduction in \( P_{\text{temporal}}(D \rightarrow E) \) compared to \( P_{\text{static}}(D \rightarrow E) \) demonstrates that static centrality measures do not necessarily translate to high influence in temporal settings. Since influence propagation in temporal networks depends on the sequence of interactions, seed nodes selected using static metrics may fail to maximize influence. Hence, a seed set optimized for a static network does not necessarily remain optimal in a temporal setting.

We have also demonstrated the variation in influence spread across different seed set sizes while maintaining a consistent seed set for both static and temporal settings. The influence propagation was analyzed using the Independent Cascade (IC) model as the diffusion mechanism, implemented via the Degree Discount heuristic \cite{murata2018extended}. The algorithm was applied to contact patterns on the Rural Malawi dataset \cite{ruralmalawi}.

\begin{figure}[ht]
    \centering
    \includegraphics[width=\linewidth]{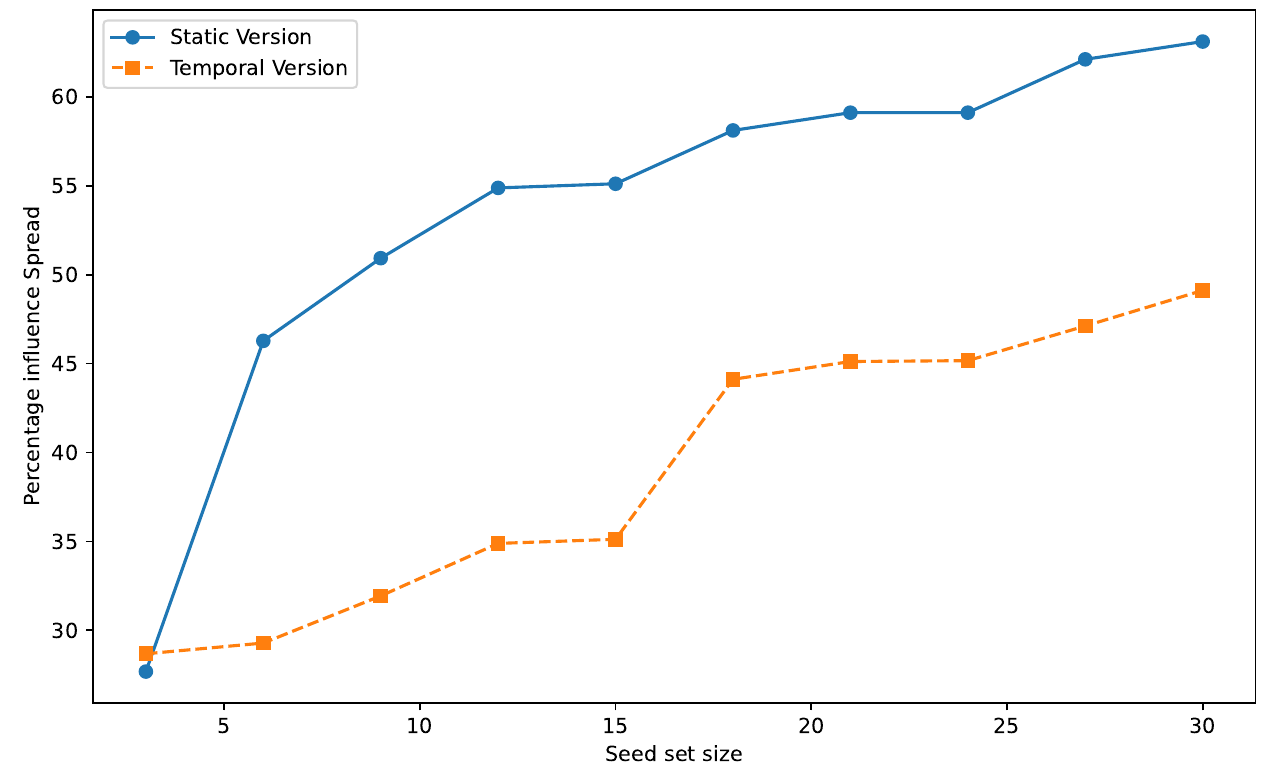}  
    \caption{Percentage influence spread on static and temporal version of Rural Malawi dataset using IC model (\( p=0.01, n=86, e=355 \)) for static case and IC model (\( p=0.01, n=86, e=102292 \)) for temporal case where number of edges include all repetitive additions across different time stamps.}
    \label{ICmodel}
\end{figure}

For the static version, the dataset was preprocessed to disregard the column containing interaction timestamps, whereas for the temporal version, interaction times were retained. Seed sets of varying sizes were determined and kept consistent across both versions. To evaluate the influence spread, we conducted 10,000 Monte Carlo simulations of the diffusion process using the IC model with a propagation probability of \( p = 0.01 \). The results, as depicted in the figure \ref{ICmodel}, illustrate the discrepancy in influence spread, underscoring that a seed set optimal for a static network is not necessarily optimal in a temporal setting.

This analysis highlights the necessity of modifying diffusion models beyond merely adapting influence maximization algorithms for their application in temporal networks. In the next section, we delve into various seed selection mechanisms that aim to identify the most influential nodes for initiating diffusion. These mechanisms are adapted to different network settings and diffusion behaviors, and understanding them is crucial for designing effective influence maximization strategies.

\section{Seed Selection Mechanisms}
Identifying influential seed nodes in temporal networks presents unique challenges due to dynamic structures, evolving diffusion patterns, and computational constraints. Several strategies have emerged to address this task, each grounded in different methodological foundations and optimization goals. Based on their core principles and the nature of adaptation to temporal dynamics, we classify seed selection mechanisms into four major categories: (1) Greedy and Heuristic Methods, (2) Dynamic Optimization Methods, (3) Incremental and Real-Time Updates, and (4) Predictive and Exploratory Techniques. A comparative summary of these approaches is presented in Table~\ref{tab:seed-selection-mechanisms}.

\subsection{Greedy and Heuristic Methods}
Early solutions to influence maximization adopted classical greedy approaches assuming static networks. These methods were extended to temporal settings with adaptations in diffusion modeling and influence estimation. For instance, CELF and CELF++ frameworks \cite{leskovec2007cost, osawa2015selecting} reduced computation by applying lazy-forward heuristics. Aggarwal et al. \cite{aggarwal2012influential} utilized influence trees for incremental estimation. More recent studies introduced entropy-based heuristics \cite{michalski2014seed, michalski2020entropy} and extended cost-aware selection under dynamic node activity \cite{murata2018extended}. While these techniques offer strong approximation guarantees, their static nature limits adaptability to evolving network structures.

\subsection{Dynamic Optimization Methods}
To cope with changes in network topology or activity over time, dynamic optimization strategies were developed. These methods aim to reuse computation by updating influence estimates in response to snapshot changes, rather than recomputing from scratch. Sketch-based optimizations \cite{song2016influential}, dynamic CELF variants \cite{ohsaka2016dynamic}, and snapshot-based forward sampling \cite{wu2019maximizing} exemplify this paradigm. Although such methods improve runtime efficiency, they incur high costs when network updates are frequent or large-scale, and still depend on accurate influence spread estimation.

\subsection{Incremental and Real-Time Updates}
A more recent trend focuses on adapting to real-time network events such as edge deletion, node churn, or topical drift. Algorithms in this class, such as StreamIM \cite{wang2017real} and DynaGraph \cite{peng2021dynamic}, track top-k influencers by maintaining subgraph updates incrementally. Topic-sensitive models \cite{min2020topic} refine seed choices based on evolving content themes. These methods offer responsiveness and low latency, making them suitable for streaming settings; however, they often compromise influence optimality due to constraints on computational overhead and the hardness of maintaining spread guarantees under continual updates.

\subsection{Predictive and Exploratory Techniques}
A forward-looking category leverages structural cues and link prediction to identify future influential nodes. Studies like \cite{linkpredim, deeplinkpred} utilize temporal link prediction and community evolution to forecast diffusion-relevant nodes, while exploratory techniques \cite{zhuang2013influence, han2017influence} emphasize information gain through structural holes and bridge nodes. Such strategies are promising for proactive influence campaigns, particularly in networks with predictable evolution. However, their effectiveness depends heavily on the accuracy of the prediction model and their robustness to uncertainty in volatile settings.

These categories are not mutually exclusive; hybrid methods often integrate predictive insight into greedy frameworks or combine real-time updating with sketch-based strategies. Despite progress, scalability, robustness to uncertainty, and temporal generalizability remain critical challenges in seed selection for temporal networks. Table~\ref{tab:seed-selection-mechanisms} provides a concise overview of each category, their representative works, and key limitations.

While we have explored various algorithms for seed selection, it is important to note that their effectiveness fundamentally depends on the underlying diffusion model used to simulate the spread. Given the vast diversity and complexity of diffusion models, it becomes essential to organize and understand them systematically. The following section presents a structured taxonomy of diffusion models to support informed model selection and accurate evaluation of seed selection strategies

\begin{table*}[htp]
    \centering
    \setlength{\belowcaptionskip}{0pt} 
    \renewcommand{\arraystretch}{1.2} 
    \resizebox{\textwidth}{!}{
    \begin{tabularx}{\textwidth}{|l|l|X|X|}
        \hline
        \textbf{Category} & \textbf{Strategy} & \textbf{Representative Works} & \textbf{Limitations} \\
        \hline
        Greedy and Heuristic Methods & Static Optimization & Greedy influence maximization using CELF \cite{leskovec2007cost} and CELF++ \cite{osawa2015selecting}; influence tree-based heuristics \cite{aggarwal2012influential}; entropy-based seed selection \cite{michalski2014seed, michalski2020entropy}; cost-aware node selection under dynamic activity \cite{murata2018extended}; heuristic evaluations in temporal influence spread \cite{erkol2020influence}. & Limited adaptability to structural or temporal changes; often ignore evolving network dynamics. \\
        \hline
        Dynamic Optimization Methods & Time-Aware Optimization & Time-adaptive heuristics for dynamic networks \cite{chen2015influential}; sketch-based influence estimation \cite{song2016influential}; dynamic versions of CELF leveraging incremental updates \cite{ohsaka2016dynamic}; snapshot-based forward sampling with lazy strategies \cite{wu2019maximizing}. & High computational cost when update frequency is high; often rely on repeated influence spread estimation. \\
        \hline
        Incremental and Real-Time Updates & Streaming and Deletion-Aware & Incremental influence tracking in evolving networks \cite{wang2017incremental}; StreamIM for real-time influence estimation \cite{wang2017real}; adaptive graph reconfiguration through DynaGraph \cite{peng2021dynamic}; content-aware topical influence tracking \cite{min2020topic}; event-driven influence adaptation \cite{chandran2022dynamic}. & May sacrifice optimality for speed; effectiveness limited by hardness of maintaining influence guarantees in dynamic settings. \\
        \hline
        Predictive and Exploratory Techniques & Forward-Looking Seed Selection & Temporal link prediction for seed forecasting \cite{linkpredim}; GNN-based prediction with dynamic features \cite{deeplinkpred}; influence forecasting via structural hole theory \cite{zhuang2013influence}; exploration-based seed selection strategies \cite{han2017influence}; graph topology-driven influence scoring \cite{zhou2009graph}. & Highly dependent on prediction accuracy and exploration strategy; susceptible to noise in fast-changing or uncertain environments. \\
        \hline
    \end{tabularx}
    }
    \caption{Summary of Seed Selection Mechanisms on Temporal Networks.}
    \label{tab:seed-selection-mechanisms}
\end{table*}

\section{Taxonomy of Diffusion Models}
The study of diffusion models is fundamental to understanding how influence, information, and behaviors spread across networks. These models provide structured frameworks for capturing real-world dynamics, from viral marketing and disease transmission to social mobilization. Given the diversity of diffusion models, each is designed to accommodate specific network structures, behavioral assumptions, and practical applications.

To build a systematic understanding, we first examine foundational models such as the Independent Cascade (IC), Linear Threshold (LT) and Suspectible Infected Recovered (SIR) frameworks, which serve as the basis for numerous extensions incorporating reinforcement mechanisms, reactivation processes, and time-sensitive adaptations.

Beyond these, we explore the extensions of standard diffusion models that capture behavioral variations, structural complexities, and external influences, enabling more precise analysis of real-world diffusion processes. These models offer deeper insights into the subtleties of influence propagation.

Finally, we present a comprehensive taxonomy classifying diffusion models based on their fundamental principles and operational mechanisms.  By structuring diffusion models in this manner, we provide a clear framework for selecting the most suitable approach, ensuring both accuracy and scalability.

\subsection{Standard Diffusion Models}  
The exploration of diffusion models begins with foundational frameworks that have become the cornerstone of influence propagation research. These standard models provide a structured and mathematically rigorous approach to understanding how information, behaviors, or contagions spread across networks. Among these, the Independent Cascade (IC) model, the Linear Threshold (LT) model, and the Susceptible-Infected-Recovered (SIR) model stand out as widely adopted paradigms. Each model captures distinct aspects of propagation dynamics, and help us dive deep and better understand different real-world scenarios. By examining these frameworks in detail, we can establish a solid foundation for understanding their strengths, limitations, and practical applications.
\subsubsection{Independent Cascade (IC) Model}  
\begin{figure}[ht]
    \centering
    \includegraphics[width=\linewidth]{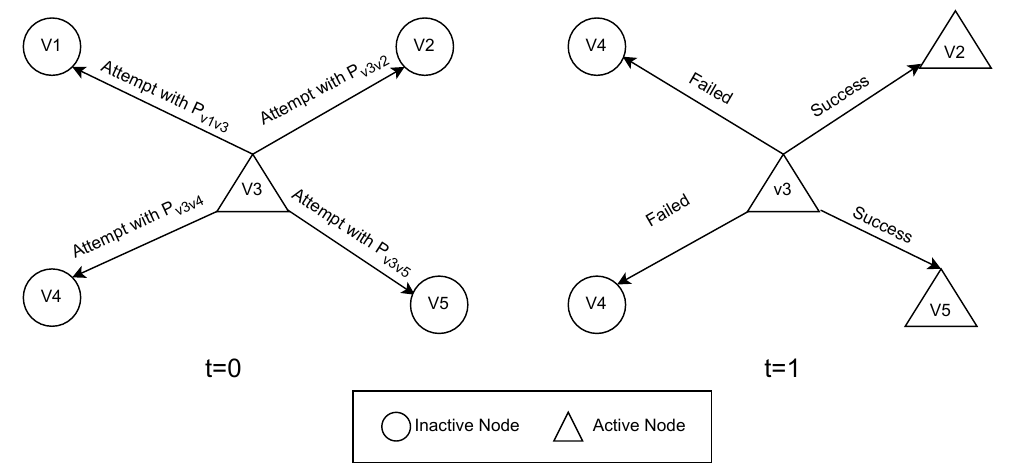}  
    \caption{Illustration of the IC model. Nodes $ V_2 $ and $ V_5 $ are successfully activated with probabilities $ P_{v_3v_2} $ and $ P_{v_3v_5} $, respectively.}
    \label{fig:ICmodel2}
\end{figure}

The IC model \cite{kempe2003maximizing} simulates influence spread using a probabilistic framework. A network is represented as a directed graph $ G = (V, E) $, where each edge $ (u, v) \in E $ has an associated probability $ p_{uv} $, denoting the likelihood of $ u $ activating $ v $. The process begins with an initial seed set $ A_0 $. At each time step $ t $, newly activated nodes attempt to influence their inactive neighbors with probability $ p_{uv} $. The process continues iteratively until no further activations occur (see figure \ref{fig:ICmodel2}).

\subsubsection{Linear Threshold (LT) Model}  
\begin{figure}[ht]
    \centering
    \includegraphics[width=\linewidth]{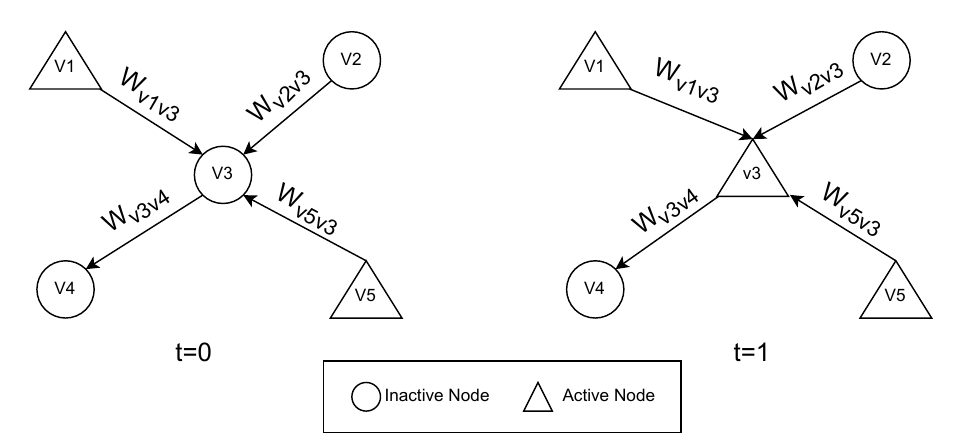}  
    \caption{Illustration of the LT model. Node $ V_3 $ becomes active if $ W_{v_1v_3} + W_{v_5v_3} \geq \theta_{v_3} $.}
    \label{fig:LTmodel}
\end{figure}

The LT model \cite{granovetter1978threshold} defines influence propagation through cumulative threshold activation. Each node $ v \in V $ has a threshold $ \theta_v $ sampled from $ [0,1] $, determining the required influence for activation. Every edge $ (u, v) \in E $ carries a weight $ w_{uv} $, ensuring $ \sum_{u} w_{uv} \leq 1 $. An inactive node $ v $ becomes active when the sum of influences from its active neighbors exceeds $ \theta_v $. The process continues until no additional activations occur (see figure \ref{fig:LTmodel}).

\subsubsection{Susceptible-Infected-Recovered (SIR) Model}  
\begin{figure}[ht]
    \centering
    \includegraphics[width=\linewidth]{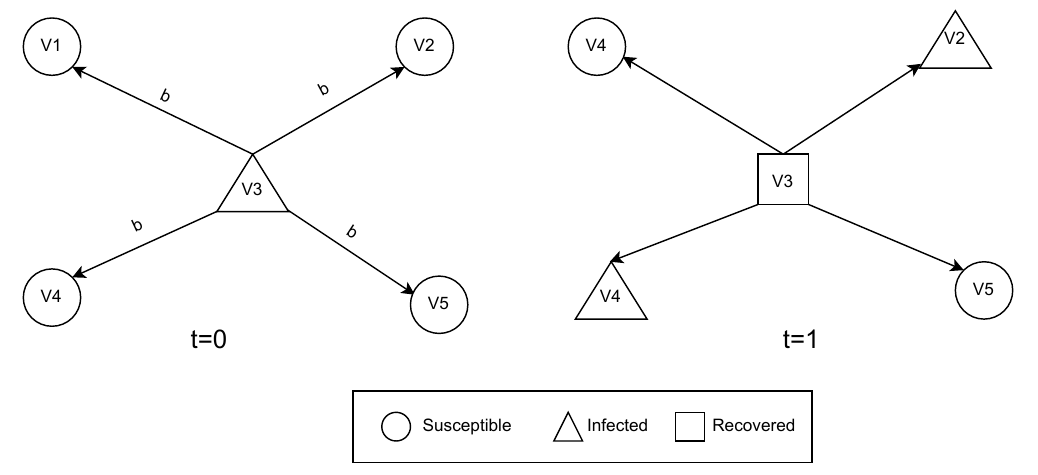}  
    \caption{Illustration of the SIR model. Node $ V_3 $ recovers at time $ t=1 $ with probability $ \gamma $, while some susceptible nodes become infected with probability $ 1-e^{-\beta k} $.}
    \label{fig:SIRmodel}
\end{figure}

The SIR model \cite{kempe2003maximizing} is widely used for modeling epidemic spread. Each node exists in one of three states: Susceptible (S), Infected (I), or Recovered (R). Transitions are governed by infection rate $ \beta $ and recovery rate $ \gamma $. Susceptible nodes become infected based on interactions with infected neighbors, with probability $ 1 - e^{-\beta k} $, where $ k $ is the number of infected contacts. Infected nodes recover with probability $ \gamma $, transitioning to the recovered state. The process follows the differential equations:

\begin{equation}
    \frac{dS}{dt} = -\beta SI, \quad \frac{dI}{dt} = \beta SI - \gamma I, \quad \frac{dR}{dt} = \gamma I
\end{equation}

until no further infections or recoveries occur (see figure \ref{fig:SIRmodel}).

While standard diffusion models have laid the foundational groundwork for understanding information spread, real-world complexities often necessitate more nuanced modelling approaches, paving the way for several important extensions of these classical models.

\subsection{Extensions}
Standard diffusion models have undergone significant refinement, incorporating diverse extensions to capture the intricacies of contagion phenomena in epidemics and other scenarios. For example, the Susceptible-Exposed-Infectious-Recovered (SEIR) framework \cite{wang2014seir} introduces an intermediate exposed phase, while the SCIR model \cite{ding2015research} integrates a tiered interaction structure. The irSIR model \cite{cannarella2014epidemiological} emphasizes recovery dynamics, whereas the ESIS model \cite{wang2015esis} accounts for emotional dimensions in transmission. Additionally, the Fractional SIR approach \cite{feng2015competing} employs fractional calculus to model memory effects in contagion processes. Fractional calculus is a generalization of standard calculus, allowing derivatives and integrals of non-integer order. This mathematical tool is particularly useful in modeling systems where past states influence current dynamics, enabling better representation of memory or history effects. A comprehensive overview of these developments can be found in \cite{li2017survey}. Murata et al. \cite{murata2018extended} adapt heuristic techniques to temporal networks to address the influence maximization (IM) problem using the Susceptible-Infectious (SI) paradigm. Among their contributions, the Dynamic Degree Discount, Dynamic CI, and Dynamic RIS methods stand out as significant enhancements. Specifically, the Dynamic Degree Discount method is tailored exclusively for the SI framework, assuming that once nodes are infected, they remain so indefinitely, a critical assumption for tracking the diminishing degree of neighboring nodes over time.

Efforts to mitigate the computational demands of greedy algorithms often involve approximations to estimate the impact of seed sets $S$. Agrawal et al. \cite{aggarwal2012influential} utilize such an approximation method within the Independent Cascade (IC) model. These approximations have been further refined using the SI framework \cite{osawa2015selecting} to improve runtime efficiency and extended to the Susceptible-Infectious-Recovered (SIR) model \cite{erkol2020influence} to assess algorithmic robustness under noisy conditions.

Zhang et al. \cite{zhang2023targeted} present a novel approach to simulate contagion propagation in networks using the SIR model. In a network comprising $N$ vertices and $M$ edges, the adjacency matrix $\{a_{ij}\}$ encodes connectivity patterns. Nodes may exist in one of three states: susceptible, infectious, or recovered. The infection probability is denoted by $\beta$, while the recovery rate is fixed at $\gamma = 1$. The system's evolution is governed by the following set of differential equations:

\begin{equation}
\frac{dS_i(t)}{dt} = -S_i(t) \left[ 1 - \prod_{j}(1 - \beta a_{ij} I_j(t)) \right],
\end{equation}
\begin{equation}
\frac{dI_i(t)}{dt} = S_i(t) \left[ 1 - \prod_{j}(1 - \beta a_{ij} I_j(t)) \right] - \frac{I_i(t)}{\gamma},
\end{equation}
\begin{equation}
\frac{dR_i(t)}{dt} = \frac{I_i(t)}{\gamma}.
\end{equation}

Building on this work, Zhang et al. \cite{zhang2024influence} extend the analysis to simplicial complexes (SCMs), which incorporate pairwise interactions (1-simplices) and higher-order collective interactions (2-simplices). Simplicial complexes provide a framework to model relationships not limited to pairs of nodes, but also groups of three or more, capturing the dynamics of collective group interactions. Here, the infection probabilities are represented by the vector $\mathbf{B} = \{\beta_1, \beta_2\}$, where $\beta_1$ governs individual links and $\beta_2$ governs group interactions. The dynamics in SCMs are described by the following equations:

\begin{equation}
\begin{aligned}
\frac{dS_i(t)}{dt} &= -S_i(t) \Bigg[ 1 - \prod_{j} \big(1 - \beta_1 \tilde{A}_{ij} I_j(t)\big) \\
&\quad \times \prod_{k,l} \big(1 - \beta_2 \tilde{B}_{ikl} I_k(t) I_l(t)\big) \Bigg]
\end{aligned}
\end{equation}

\begin{equation}
\begin{aligned}
\frac{dI_i(t)}{dt} &= S_i(t) \Bigg[ 1 - \prod_{j} \big(1 - \beta_1 \tilde{A}_{ij} I_j(t)\big) \\
&\quad \times \prod_{k,l} \big(1 - \beta_2 \tilde{B}_{ikl} I_k(t) I_l(t)\big) \Bigg] - \frac{I_i(t)}{\gamma}
\end{aligned}
\end{equation}

\begin{equation}
\frac{dR_i(t)}{dt} = \frac{I_i(t)}{\gamma}.
\end{equation}

Their study centers on optimizing initial seed selection to maximize influence among target nodes ($V_T$) while minimizing it among non-target nodes ($V_{NT}$). The optimal seed configuration minimizes the ratio $\frac{g(s_{n^*})}{f(s_{n^*})}$ for $f(s_{n^*}) > 0$, where $f(s_{n^*}) = \lim_{t \to \infty} \sum_{i \in V_T} R_i(t)$ and $g(s_{n^*}) = \lim_{t \to \infty} \sum_{i \in V_{NT}} R_i(t)$.

In a related contribution, Liu et al. \cite{liu2014information} propose an SIR model with a fixed recovery period to analyze information dissemination. Individuals are classified into three categories: uninformed (S-state), actively informed (I-state), and inactive informed (R-state). The process begins with a randomly chosen node in the I-state, spreading information to its S neighbors before transitioning to the R-state post-transmission. The propagation ceases when no active informants remain. To enhance reach, a rewiring mechanism allows I-state nodes to redirect connections to randomly selected S-state nodes among their second-order neighbors, guided by probabilities derived from the Fermi function. The Fermi function is a probabilistic function commonly used in statistical physics and evolutionary game theory to model stochastic decisions, capturing how likely a node is to adopt a behavior based on the perceived benefit or influence strength. Traditional diffusion models also often assume a one-time activation of nodes, overlooking the crucial aspect of active-inactive transitions. Zahoor et al. \cite{zahoor2024influencemaximizationtemporalnetworks} proposed Continuous Persistent Susceptible-Infected Model with Reinforcement and Re-activation (cpSI-R) which addresses this limitation by explicitly incorporating these transitions. By allowing nodes to regain influence potency through reactivation and reinforcement, cpSI-R provides a more realistic representation of diffusion processes. This approach ensures that influence accumulation is not underestimated, leading to a more accurate estimation of influence spread. Moreover, the model's monotonic and submodular properties enable efficient optimization for seed selection, aligning well with theoretical guarantees in influence maximization.

While these standard models and their extensions provide foundational insights, they represent only a subset of diffusion frameworks. Advanced models incorporate behavioral variability, structural complexity, and external factors, enhancing the accuracy of diffusion analysis. To systematically categorize these models, we introduce a taxonomy based on underlying objectives and mechanisms. This classification includes process-oriented, interaction-oriented, competition-oriented, structure-oriented, and target-oriented models, each addressing distinct aspects of influence propagation. This classification offers researchers and practitioners a structured approach to selecting the most appropriate diffusion model for their applications, balancing accuracy and computational efficiency. Our taxonomy places particular emphasis on models designed for temporal networks.

\subsection{Process Oriented Models}
Process-oriented models focus on the stepwise evolution of information diffusion, capturing how activation propagates through structured mechanisms. We classify these models into three categories: explanatory, epidemic and predictive models. Predictive models are further classified into threshold, and cascading models, each addressing distinct aspects of information propagation.

We have already discussed explanatory models, which provide insights into the underlying mechanisms of diffusion, and epidemic models, which are rooted in epidemiology and simulate contagion-like spread. Predictive models aim to forecast the future state of information diffusion based on historical observations and network evolution and form the broader class of threshold and cascading models.

\subsubsection{Threshold Models}

Threshold-based models provide a fascinating lens through which to examine collective behavior, leveraging threshold values to delineate decision-making boundaries at both individual and group levels. Originally conceptualized by Mark Granovetter \cite{granovetter1978threshold}, these models have been instrumental in elucidating phenomena such as the diffusion of innovations and shifts in public opinion. By considering thresholds influenced by factors like socioeconomic status and education, the models reveal how individual choices collectively shape group dynamics. Remarkably, even minor adjustments to these thresholds can trigger profound transformations in collective actions, underscoring the intricate balance between personal utility and social influence. This can be seen in classical variants, hybrid models and temporal models as discussed below.

\paragraph{Classical Threshold Variants}

The diversity of threshold models is vast, each variant offering unique insights tailored to specific applications. The Majority Threshold Model, for instance, activates a node when the majority of its neighbors are active, making it particularly relevant for systems like voting mechanisms and distributed computing \cite{peleg1998size, peleg2002local}. Despite its simplicity, solving the influence maximization problem within this framework remains computationally challenging, akin to the general case \cite{chen2009approximability}.

Similarly, the Small Threshold Model, with thresholds as low as $\theta_v = 1$ or $2$, simplifies certain scenarios but retains its NP-hard complexity for higher thresholds, highlighting the persistent computational intricacies of even seemingly straightforward configurations.

On the other hand, the Unanimous Threshold Model imposes stringent conditions for activation, requiring all neighbors to be active before a node transitions. This model's resilience to influence makes it highly applicable in domains such as network security and epidemic containment \cite{chen2009approximability}.

\paragraph{Hybrid Threshold Models}

Beyond their foundational forms, threshold models have been extended to incorporate more nuanced dynamics. For example, the Linear Threshold with Color (LT-C) model integrates user experience and product adoption into the activation process, moving beyond traditional influence metrics \cite{bhagat2012maximizing}.

Other adaptations allow nodes to toggle between active and inactive states \cite{pathak2010generalized}, or enable simultaneous diffusion across multiple interconnected networks \cite{nguyen2013influence, shen2012interest}. The Decaying Reinforced User-centric (DRUC) model, introduced by Lagnier et al. \cite{lagnier2013predicting}, combines user profiles with information content, factoring in user intent, interest, and neighbor influence to model diffusion more realistically.

Chen and Yitong \cite{chen2012threshold} proposed a heuristic algorithm that dynamically computes Potential Influence Nodes (PIN) for optimized seed selection. Chen et al. \cite{chen2010scalable} introduced the LDA G algorithm, enabling scalable influence maximization on networks with millions of nodes and edges. Pathak et al. \cite{pathak2010generalized} extended the linear threshold model using rapidly mixing Markov chains to simulate varied cascade behaviors. Zhang et al. \cite{zhang2014recent} presented an algorithm to estimate the most probable cascade spread, demonstrating effectiveness on global-scale datasets.

He et al. \cite{he2012influence} introduced the Competitive Linear Threshold (CLT) model to address competitive influence spread, focusing on strategic seed selection to counteract rival effects. Litou et al. \cite{litou2016real} formulated the Dynamic Linear Threshold (DLT) model to combat misinformation by strategically disseminating reliable information, framing the challenge as an optimization problem to identify optimal user subsets.

\paragraph{Threshold Models in Temporal Graphs}

Dynamic graph models represent another frontier in threshold-based models. Gayraud et al. \cite{gayraud2015diffusion} introduced the Evolving Linear Threshold (ELT) model, comprising two variants: the Transient ELT (tELT) and Persistent ELT (pELT). In the tELT model, a node's activation depends on the cumulative influence of its active neighbors within a single snapshot, capturing ephemeral diffusion processes. Conversely, the pELT model allows nodes to accumulate influence over time from all active neighbors encountered, reflecting persistent influence buildup. Notably, the pELT model exhibits monotonicity and submodularity, ensuring that influence accumulates progressively, with optimal activation timing occurring early in the process.

These developments underscore the versatility of threshold models in addressing real-world challenges, employing sophisticated algorithms to pinpoint the most impactful seed nodes for influence propagation. Through their ability to capture complex interplays between individual decisions and collective outcomes, threshold models continue to illuminate the mechanisms driving influence spread and maximization. Their adaptability and robustness make them indispensable tools for understanding and shaping dynamic systems across diverse domains.

\subsubsection{Cascading Models}

Cascading models, rooted in principles from particle systems and probability theory \cite{liggett1985interacting} \cite{durrett1988lecture}, serve as powerful tools for understanding diffusion processes across diverse domains, including marketing strategies and social network dynamics \cite{goldenberg2001talk} \cite{goldenberg2001using}. While the Independent Cascade (IC) model has been briefly introduced earlier, its pivotal role in solving the Influence Maximization (IM) problem cannot be overstated. Over the years, researchers have refined and extended this model to address challenges such as scalability, opinion dynamics and real-world applicability.

\paragraph{Model Extensions and Scalability Enhancements}

To estimate propagation probabilities within the IC framework, Saito et al. \cite{saito2008prediction} employed an Expectation-Maximization (EM) algorithm. However, the computational overhead of this approach limits its feasibility for large-scale networks. Addressing scalability, Wang et al. \cite{wang2012scalable} and Jung et al. \cite{jung2012irie} emphasized the need for efficient algorithms. Arora et al. \cite{galhotra2015asim} developed the ASIM algorithm, which balances runtime efficiency and memory usage, making IC more practical for real-world networks. Further, Barbieri et al. \cite{barbieri2013topic} proposed topic-aware variants, TIC and TLT, which consider the thematic relevance of propagated content, enhancing both scalability and contextual applicability.

\paragraph{Incorporating Opinion Dynamics }

Chen et al. \cite{chen2011influence} extended the IC model to include negative opinions in the diffusion process. A quality factor $q$ determines whether a newly activated node adopts a positive or negative stance, reflecting psychological dynamics like negativity bias and dominance \cite{rozin2001negativity}. Nodes adopting a negative stance remain so in all subsequent rounds. An efficient influence computation algorithm was also proposed for tree structures, forming the basis for heuristics on general graphs. In another development, the Decreasing Cascading Model (DC) modifies the IC framework by modeling diminishing activation probabilities \cite{kempe2005influential}, which captures the saturation effect of repeated exposure.

\paragraph{Temporal Variants}

Further adaptations address temporal realism. Kim et al. \cite{kim2014ct} introduced the CT-IC model, relaxing the single-attempt activation assumption; active nodes repeatedly attempt to activate their neighbors within a limited timeframe. Similarly, Zhu et al. \cite{zhu2014maximizing} proposed the Continuous Time Markov Chain model (CTMC-ICM), optimizing influence spread by identifying influential node subsets. Zhang et al. \cite{zhang2014recent} advanced a generalized cascade model where activation probabilities depend on the set of influencing neighbors, preserving order-independence and equivalence to generalized threshold models.

Yang et al. \cite{yang2019influence} introduced the t-IC model, which includes constraints such as activation timing and repetition, offering better realism for temporal networks. Complementing this, the Evolving Independent Cascade (EIC) model by Gayraud et al. comprises two variants: Transient EIC (tEIC) and Persistent EIC (pEIC). The tEIC assumes immediate activation, suitable for scenarios like disease transmission, while the pEIC supports long-term influence propagation, such as product adoption. Importantly, the spread function in pEIC is monotone and submodular under constant activation probabilities, aligning well with optimization strategies in evolving networks.

\paragraph{Memory and Historical Behavior in Diffusion}

Hao et al. \cite{hao2011influence} introduced the Time-Dependent Comprehensive Cascade (TCC) model, which integrates historical activations. Each node at time $t$ makes a single activation attempt, with the probability influenced by previous failures. The activation probability $p_{u,t,v}$ depends on $S_t$, the set of neighbors who previously failed to activate $v$. A parameter $K$ modulates the transition: $K = -1$ increases activation probability, $K = 1$ decreases it, and $K = 0$ retains the original probability. When $K = 1$, the TCC model reduces to the standard IC model.

Aggarwal et al. \cite{aggarwal2012influential} proposed a probabilistic variant of the IC model to account for information possession over time. The probability that node $i$ holds information at time $t_2$ is:
\begin{equation}
\begin{aligned}
\pi(i,t_2) &= \pi(i,t_1) + (1 - \pi(i,t_1)) \\
&\quad \times \big(1 - \text{p}(\text{no transmission from neighbors})\big),
\end{aligned}
\end{equation}
where $\pi(i,t)$ represents the probability of node $i$ being informed at time $t$, and ``p" denotes the probability of zero transmissions from neighbors in the interval $[t_1, t_2]$.

These advancements highlight the adaptability of cascading models in capturing complex influence dynamics, from scalability and temporal effects to psychological and memory-aware behaviours. They offer a better understanding of the underlying diffusion in complex systems.

The models are further summarized in table \ref{processoriented}

\begin{table*}[ht] 
    \centering
    \renewcommand{\arraystretch}{0.95}
    \small
    \resizebox{\textwidth}{!}{%
    \begin{tabular}{|p{3.5cm}|p{3cm}|p{2cm}|p{2cm}|p{2.5cm}|}
    \hline
    \textbf{Diffusion Model} & \textbf{Network Type} & \textbf{Submodular} & \textbf{Monotone} & \textbf{Type} \\
    \hline
    IC~\cite{kempe2003maximizing} & Static & $\checkmark$ &$\checkmark$&Predictive \\
    SI~\cite{wang2012scalable} & Static and Temporal & $\checkmark$ &$\checkmark$&Epidemic \\
    SIR~\cite{kempe2003maximizing} & Static and Temporal & $\checkmark$ &$\checkmark$&Epidemic \\
    SIS~\cite{durrett1988lecture} & Static & $\checkmark$ &$\checkmark$&Epidemic \\
    SCIR~\cite{ding2015research} & Static & $\checkmark$ &$\checkmark$&Explanatory \\
    irSIR~\cite{cannarella2014epidemiological} & Static & $\checkmark$ &$\checkmark$&Explanatory \\
    FSIR~\cite{feng2015competing} & Static & $\checkmark$ &$\checkmark$&Explanatory \\
    SEIR~\cite{wang2014seir} & Static & $\checkmark$ &$\checkmark$&Explanatory \\
    SCM~\cite{zhang2024influence} & Static and Temporal & $\times$ &$\times$&Explanatory \\
    ESIS~\cite{wang2015esis} & Static & $\checkmark$ &$\checkmark$&Predictive \\
    cpSI-R~\cite{zahoor2024influencemaximizationtemporalnetworks} & Temporal & $\checkmark$ &$\checkmark$&Explanatory \\
    LT~\cite{kempe2003maximizing} & Static and Temporal & $\times$ &$\times$&Threshold \\
    MTM~\cite{peleg1998size} & Static & $\checkmark$ &$\checkmark$&Threshold \\
    STM~\cite{peleg2002local} & Static & $\checkmark$ &$\checkmark$&Threshold \\
    UTM~\cite{chen2009approximability} & Static & $\checkmark$ &$\checkmark$&Threshold \\
    OCM~\cite{chen2012threshold} & Static & $\checkmark$ &$\checkmark$&Threshold \\
    LTC~\cite{bhagat2012maximizing} & Static & $\checkmark$ &$\checkmark$&Threshold \\
    GTM~\cite{pathak2010generalized} & Static & $\checkmark$ &$\checkmark$&Threshold \\
    DLT~\cite{litou2016real} & Temporal & $\times$ &$\times$&Threshold \\
    tELT~\cite{gayraud2015diffusion} & Temporal & $\checkmark$ &$\checkmark$&Threshold \\
    pELT~\cite{gayraud2015diffusion} & Temporal & $\checkmark$ &$\checkmark$&Threshold \\
    CLT~\cite{he2012influence} & Static & $\checkmark$ &$\checkmark$&Threshold \\
    DRUC~\cite{lagnier2013predicting} & Static & $\checkmark$ &$\checkmark$&Threshold \\
    TBasic~\cite{yang2019influence} & Temporal & $\times$ &$\times$&Cascading \\
    ASIC~\cite{galhotra2015asim} & Static & $\checkmark$ &$\checkmark$&Cascading \\
    ASLT~\cite{barbieri2013topic} & Static & $\times$ &$\times$&Threshold \\
    TCC~\cite{hao2011influence} & Temporal & $\times$ &$\times$&Cascading \\
    CTM-IC~\cite{zhu2014maximizing} & Temporal & $\times$ &$\times$&Cascading \\
    \hline
    \end{tabular}%
    }
    \caption{Process-Oriented Diffusion Models}
    \label{processoriented}
\end{table*}

\subsection{Interaction-Oriented Models}

Interaction-driven models provide a systematic framework to guide how interactions drive
the spread of information and behaviours. . These models are categorized into two primary types: pairwise interaction models and group-level interaction frameworks. While pairwise models emphasize direct connections between individuals, group-level models explore collective dynamics within clusters or communities, shedding light on how social structures amplify influence diffusion. 

\subsubsection{Pairwise Interaction Models}Pairwise interaction models play a crucial role in social network analysis, focusing on key influencers who act as catalysts in the diffusion of information. Identifying these influencers involves leveraging network structure, mutual information, and user attributes, often employing heuristic algorithms to detect and rank influential nodes.

One foundational model in this domain is the voter model proposed by Holley et al. \cite{holley1975ergodic}, where each node in a network probabilistically adopts a neighbor’s opinion. 

Specifically, a node switches to a neighbor’s opinion with probability $p$ or retains its current opinion with probability $1 - p$. The evolution of opinions follows the equation:
\begin{equation}
P_{t+1}(i, j) = p \cdot \frac{k_j}{2m} P_t(j, i) + (1-p) \delta_{ij}
\end{equation}

where $k_j$ denotes the degree of node $j$, $m$ represents the total number of edges, and $\delta_{ij}$ is the Kronecker delta.

Building on this, Gastner et al. \cite{gastner2019voter} extend the voter model (EVM) to networks with exogenous community structures. In this framework, a network is partitioned into two cliques connected by $X$ inter-clique edges, with each node adopting either a red or blue opinion. Opinion updates occur at exponential intervals, forming a continuous-time Markov chain. To simplify the analysis, the system is represented using macrostates $(\rho_1, \rho_2)$, where $\rho_i$ denotes the fraction of red agents in clique $i$. The transition rate equation is given by:
\begin{equation}
\begin{aligned}
Q_{(\rho_1, \rho_2), (\rho_1 + 1/\alpha N, \rho_2)} &= \alpha N(1 - \rho_1)  \cdot \left( \frac{\alpha N - 1}{\alpha N} \right) \rho_1 \\
&\quad + \frac{X}{\alpha N} \cdot \frac{\rho_2}{\alpha N - 1} + \frac{X}{\alpha N}.
\end{aligned}
\end{equation}
This equation quantifies the rate at which the fraction of red agents in clique 1 increases by $1/\alpha N$ while maintaining clique 2’s fraction unchanged. In each iteration, a node is randomly selected and adopts the opinion of a randomly chosen neighbor, facilitating opinion spread dynamics across communities.

Berenbrink et al. \cite{berenbrink2016bounds} extend this analysis to dynamic networks by introducing the dynamic voter model (DVM), where nodes randomly select neighbours  and update their opinions at each time step. In dynamic graphs with conductance at least $\phi$, where edges are rewired in every round,the expected consensus time is given by $O\left(\frac{m}{d_{\text{min}} \cdot \phi}\right)$. Moreover, in the biased dynamic voter model (BDVM), nodes adopt the opinion with the highest probability when there is a substantial gap between the top two opinion probabilities. In this scenario, convergence time is bounded by $O(\log n / \phi)$.

Acemoglu et al. \cite{acemoglu2014dynamics} introduce the Information Exchange Model (IEM), where agents base decisions on private signals \( s_i \) and messages received from neighbors \( m_{ji} \) through a communication network \( G_n \). Each agent selects an action \( \sigma_{i,t} \) at time \( t \), optimizing an expected payoff function:
\begin{equation}
U_t = \max \left\{ \mathbb{E} \left[\psi - (x - \theta)^2 \mid I_t \right], \lim_{dt \to 0} e^{-r dt} \mathbb{E} \left[ U_{t+dt} \mid I_t \right] \right\}.
\end{equation}

This equation represents a tradeoff between acting immediately based on current information \( I_t \), and deferring the decision to a future time \( t + dt \) in anticipation of gaining better information. The term \( \mathbb{E}[\psi - (x - \theta)^2 \mid I_t] \) reflects the expected immediate utility from choosing action \( x \), while the discounted future utility term accounts for the value of waiting, with \( r \) denoting the discount rate. This formulation captures how agents balance learning and action in dynamic environments. Here, strategic information exchange influences network dynamics as agents modify their actions based on evolving information sets. The model examines the timing of irreversible decisions, ensuring optimal decision-making under uncertainty.

Several studies have investigated methodologies for identifying influential nodes and analyzing opinion dynamics. Bo et al. \cite{chen2014identifying} classified users into ordinary, active, subject opinion leaders, and network leaders based on behavior analysis. Jiaxin et al. \cite{mao2014social} predicted retweet potential by analyzing network structure and user activity. Xianhui et al. \cite{wu2015mining} developed Topic-Leader Rank (TLRA), incorporating user relationships, content engagement, and activity levels to mine topic-specific influencers. Ullah et al. \cite{ullah2017identification} designed an influence maximization model, identifying nodes that maximize information diffusion while minimizing contagion time.

Extending beyond static pairwise interactions, Chu et al. \cite{chuporter} introduced models for temporal networks, capturing non-Poisson interaction dynamics. Their approach generalizes voter models to non-Markovian settings by incorporating arbitrary waiting-time distributions (OM-WTDs). They demonstrate that models with heavy-tailed waiting times slow down opinion convergence and that agents with longer waiting times exert disproportionate influence on collective opinions.

Jain et al. \cite{Trust-and-reputation-based} further refined this framework by integrating trust and reputation scores into opinion models. Their approach considers edge credibility, dynamically evolving over time, affecting both opinion propagation and convergence behavior in scale-free networks.

Pairwise interaction models serve as fundamental building blocks for understanding influence propagation, consensus formation, and strategic decision-making in social networks. By integrating network topology, temporal dynamics, and agent-based decision frameworks, these models provide deep insights into how opinions spread and evolve in complex, dynamic environments. Future research can further explore how multi-agent reinforcement learning and game-theoretic approaches enhance our understanding of network-driven opinion formation.

\subsubsection{Group-Oriented Models}

Group-oriented models focus on the critical role of social structures and collective interactions in facilitating influence diffusion within communities. Communities, defined by shared interests or attributes, serve as fundamental building blocks in social networks. Detecting influential communities is a complex task often addressed by integrating link structures with content attributes. These models employ probabilistic frameworks, distance-based metrics, and hybrid methodologies to optimize community identification while balancing computational efficiency and accuracy.

Among the prominent approaches are PCL-DC and SA-Cluster-Inc, which leverage probabilistic models and distance calculations to refine community detection \cite{yang2014combining, zhou2010clustering}. Similarly, CODICIL and CESNA integrate link strength and content similarity to enhance detection performance \cite{ruan2013efficient, yang2013community}. Sentiment-based models and K-core algorithms further enrich community extraction by incorporating sentiment analysis and structural representations \cite{yang2014community, peng2014accelerating}. However, concerns about the subjectivity of sentiment-based methods have led to innovations like the SVO method, which improves objectivity and computational scalability \cite{gurini2015analysis}. These techniques iteratively update network structures based on content and attributes, ensuring precision without overwhelming computational demands \cite{ullah2017community}.

Anderson et al. \cite{anderson2015global} examined global diffusion patterns in LinkedIn's signup cascades, revealing homophily at both local and global scales. This study underscores how coherent member groups emerge through cascading processes across multiple timescales. Liu et al. \cite{liu2016comparing} introduced the Topic Adoption Model (TAM), which analyzes hashtag propagation to model information adoption dynamics. TAM constructs graphs based on hashtag-sharing behaviors, quantifying rational probability scores $P_r(u_i \rightarrow u_j)$ between users using the equation:
\begin{equation}
    P_r(u_i \rightarrow u_j) = \sum_{p=(u_i,u_j)} \rho(p|r),
\end{equation}
where $\rho$ represents the random walk probability of path $p$ under relationship $r$. This framework highlights how different factors contribute to topic adoption across platforms.

Fa et al. \cite{fan2014individual} proposed a preference-based diffusion model where individual acceptance depends on personal preferences and peer influence. Initially, decisions are driven by personal inclinations, but over time, influence from active friends becomes increasingly significant. The model accounts for scenarios where individuals accept information contrary to their preferences due to group influence or self-initiated adoption.

Xiong et al. \cite{xiong2015information} investigated information diffusion in microblogging platforms like Twitter and Sina Weibo, introducing a model that categorizes users into spreaders, ignorants, and terminators. Information spreads with probabilities such as the likelihood of a follower receiving a story ($\gamma = p(m) \times f_{page}(m)$), a spreader becoming a terminator ($\theta$), and a spreader continuing to propagate the story ($\alpha = e^{-t}$). The model emphasizes community structures, where information initially propagates within communities before crossing boundaries, influenced by browsing behaviors and story visibility.

Myers et al. \cite{myers2012information} developed an information diffusion model that integrates internal network-edge propagation and external influences. Contagions, representing specific pieces of information, spread as independent events. Nodes become infected upon first exposure, influenced by external exposures described by $\lambda_{ext}(t)$ and internal exposures modeled by the hazard function $\lambda_{int}(t)$.

Sun et al. \cite{SUN20141} introduced an incremental density based algorithm (IncOrder) for detecting overlapping communities in dynamic networks. Their approach combines degree-based seed selection with a cascade information diffusion model to update communities incrementally. Sattari et al. \cite{sattari2018cascade} enhanced community detection accuracy by integrating label propagation and cascade diffusion models, addressing challenges posed by new nodes joining existing communities. Their experiments on real and synthetic networks demonstrate improved detection performance.

Cui et al. \cite{cui2017modeling} explored the impact of time-varying modular structures on information dissemination in temporal networks. They introduced a continuous-time Markov model incorporating mobility rate and community attractiveness parameters. Numerical results reveal that variations in social mobility and community attractiveness significantly influence diffusion dynamics, enhancing spreading efficiency. These findings underscore the importance of considering temporal community structures in influence maximization strategies. The interaction oriented models are further summarised in Table \ref{interaction_models}.

\begin{table*}[ht]
    \centering
    \renewcommand{\arraystretch}{0.95}
    \small
    \resizebox{\textwidth}{!}{%
    \begin{tabular}{|p{5.5cm}|p{2.5cm}|p{2.5cm}|p{1.5cm}|p{1.5cm}|}
    \hline
    \textbf{Diffusion Model} & \textbf{Network Type} & \textbf{Type} & \textbf{Submodular} & \textbf{Monotone} \\
    \hline
    Voter Model \cite{holley1975ergodic} & Static & Pairwise & $\checkmark$ & $\checkmark$ \\
    Extended Voter Model (EVM) \cite{gastner2019voter} & Static & Pairwise & $\checkmark$ & $\checkmark$ \\
    Dynamic Voter Model (DVM) \cite{berenbrink2016bounds} & Static and Temporal & Pairwise & $\times$ & $\times$ \\
    Biased Dynamic Voter Model (BDVM) \cite{berenbrink2016bounds} & Static and Temporal & Pairwise & $\times$ & $\times$ \\
    Information Exchange Model (IEM) \cite{acemoglu2014dynamics} & Static & Pairwise & $\checkmark$ & $\checkmark$ \\
    Topic-Leader Rank (TLRA) \cite{wu2015mining} & Static & Group-Oriented & $\checkmark$ & $\checkmark$ \\
    Trust and Reputation Model \cite{Trust-and-reputation-based} & Static & Group-Oriented & $\checkmark$ & $\checkmark$ \\
    Opinion Model \cite{ullah2017identification} & Static & Group-Oriented & $\checkmark$ & $\checkmark$ \\
    OM-WTD Model \cite{chuporter} & Temporal & Pairwise & $\times$ & $\times$ \\
    Leader Influence Maximization (LIM) \cite{chen2014identifying} & Static and Temporal & Pairwise & $\times$ & $\times$ \\
    Preference-Oriented Exposure (POE) Model \cite{fan2014individual} & Static and Temporal & Pairwise & $\times$ & $\times$ \\
    Topic Adoption Model (TAM) \cite{liu2016comparing} & Static & Group-Oriented & $\checkmark$ & $\checkmark$ \\
    PCL-DC \cite{yang2014combining} & Static & Group-Oriented & $\checkmark$ & $\checkmark$ \\
    SA-Cluster-Inc \cite{zhou2010clustering} & Static & Group-Oriented & $\checkmark$ & $\checkmark$ \\
    CODICIL \cite{ruan2013efficient} & Static & Group-Oriented & $\checkmark$ & $\checkmark$ \\
    CESNA \cite{yang2013community} & Static & Group-Oriented & $\checkmark$ & $\checkmark$ \\
    SVO \cite{gurini2015analysis} & Static & Group-Oriented & $\checkmark$ & $\checkmark$ \\
    IncOrder \cite{SUN20141} & Static & Group-Oriented & $\times$ & $\times$ \\
    Label Propagation and Cascade Model \cite{sattari2018cascade} & Static & Group-Oriented & $\times$ & $\times$ \\
    CTMM (Continuous-Time Markov Model) \cite{cui2017modeling} & Static & Pairwise & $\times$ & $\times$ \\
    \hline
    \end{tabular}%
    }
    \caption{Interaction-Oriented Diffusion Models}
    \label{interaction_models}
\end{table*}

\subsection{Competition-Oriented Models}

Competition-oriented diffusion models examine the simultaneous spread of multiple innovations within a social network, where individuals adopt only one of the competing alternatives. These models capture the strategic decisions of firms operating under budget constraints while targeting specific consumers to maximize adoption \cite{carnes2007maximizing}. Various frameworks have been proposed to model competition-driven diffusion processes, each addressing distinct aspects of influence propagation dynamics.

Traditional models primarily focus on single-cascade diffusion, but competitive settings introduce added complexity due to the interplay between multiple influence sources. Eiselt et al. \cite{eiselt1989competitive} introduce two fundamental models for competitive diffusion. The distance-based model (DBM) aligns with competitive facility location theory, where a node's proximity to an initial adopter influences adoption likelihood. In contrast, the wave propagation model (WPM) conceptualizes diffusion as a stepwise process, wherein nodes adopt a technology based on their spatial proximity to already influenced individuals. Both approaches employ heuristic optimization techniques, such as the Hill Climbing Algorithm, to enhance adoption outcomes.

Borodin et al. \cite{borodin2010threshold} extend these ideas with the weight-proportional threshold model (WPTM) and the separated threshold model (STM). These frameworks account for real-world competitive scenarios by considering heterogeneous adoption thresholds and varying influence strengths across network edges. In these models, inactive nodes transition to an active state when their cumulative exposure surpasses a predefined threshold. The STM further refines this by assigning independent adoption thresholds for each competing alternative.

Beyond conventional diffusion, Li et al. \cite{li2013influence} explore influence propagation in online social networks (OSNs) with both positive and negative relationships, leveraging signed graphs. By extending the voter model, they analyze influence competition dynamics and propose efficient seed selection algorithms for IM problem. Their approach enhances predictive accuracy compared to prior methodologies.

Bozorgi et al. \cite{bozorgi2017community} introduce the DCM (Decision-aware Competitive Model), an enhancement of the Linear Threshold (LT) model designed to capture the deliberation process preceding adoption. The model introduces an intermediate “thinking” state where nodes evaluate competing influences before making a decision. Their findings establish the NP-hardness of competitive influence maximization under DCM, leading to the development of the CI2 algorithm, which efficiently identifies influential nodes. Compared to WPTM and STM, DCM provides a more realistic representation of competitive influence propagation.

Yu et al. \cite{yu2017fair} propose the Timeliness Independent Cascade (TIC) model to address multi-influence competition, where nodes accumulate exposure to different influence sources before making a final adoption decision. They formulate the FairInf problem, aiming to optimize seed selection for multiple firms while maintaining equitable influence distribution. Their model reflects real-world scenarios where users postpone decisions until they receive sufficient exposure to competing alternatives.

Chakraborty et al. \cite{chakraborty2019competitive} investigate budget-constrained influence maximization in competitive settings, incorporating dynamic node states and continuous resource expenditures. Using voting dynamics, they develop optimal influence strategies under both known and adversarial conditions, deriving equilibrium strategies within star-topology networks. Their findings highlight the impact of varying cost structures on influence propagation efficiency.

Gao et al. \cite{gao2020fair} extend competitive influence modeling to event-based social networks (EBSNs) by introducing the Fair-aware Competitive Event Influence Maximization (FCE-IM) framework. They propose the E-LT model, an adaptation of the LT model tailored to competitive offline social events. The model considers distinct seed sets for competing events, with activation probabilities based on cumulative influence from previously activated neighbors. A fairness-aware selection strategy ensures balanced participation across competing events.

Liu et al. \cite{liu2020algorithm} address computational efficiency in competitive influence maximization by formulating the Influence Maximization with Limited Unwanted Users (IML) problem under the Independent Cascade model. Instead of traditional influence simulation methods, they introduce a propagation path-based estimation technique that significantly reduces computational complexity. Their greedy selection algorithm achieves superior performance while minimizing runtime overhead.

Tsaras et al. \cite{tsaras2021collective} propose the Awareness-to-Influence (AtI) model, a two-phase diffusion framework that differentiates between information exposure and final adoption. In the awareness phase, nodes receive influence from multiple sources, with propagation probabilities dependent on edge weights and similarity measures. The influence phase determines the final adoption based on accumulated awareness, ensuring disjoint influence sets. Their model outperforms conventional approaches by accurately capturing real-world decision-making patterns.

Wang et al. \cite{wang2021maximizing} introduce the TrCID model, which incorporates both positive and negative influence dynamics through trust and distrust relationships. This extension of the LT model categorizes nodes into positively activated, negatively activated, or inactive states, with influence propagation governed by weighted trust values. A time-decay function ensures realistic modeling of influence attenuation over time. Their framework utilizes a flow-based trust estimation method, enabling efficient computation of trust-aware influence propagation.

Liang et al. \cite{liang2023targeted} present the Targeted Influence Competition Cascade (TICC) model, integrating product competitiveness and user specificity into influence propagation. Their approach eliminates the need for explicit competing nodes, instead modeling global competition strength via a competition coefficient. The propagation process ensures that activation decisions occur in a structured manner, reflecting real-world competitive marketing strategies. Zahoor et al. \cite{tbcelf} offers a two-fold solution for competitive marketing. Firstly, it optimizes temporal seed selection, extending the principles of cost-effective lazy forward optimization. Secondly, it imposes a budget constraint, ensuring efficient seed selection within budgetary limits.

The models are further summarized in table \ref{competition_models}.

\begin{table}[ht]
    \centering
    \renewcommand{\arraystretch}{1.1}
    \footnotesize
    \begin{tabular}{|p{3.7cm}|p{1cm}|p{1.2cm}|p{1.2cm}|}
    \hline
    \textbf{Diffusion Model} & \textbf{Network Type} & \textbf{Submodular} & \textbf{Monotone} \\
    \hline
    Distance-Based Model (DBM)~\cite{eiselt1989competitive} & Static & $\checkmark$ & $\times$ \\
    Wave Propagation Model (WPM)~\cite{eiselt1989competitive} & Static & $\checkmark$ & $\times$ \\
    Weight-Proportional Threshold Model (WPTM)~\cite{borodin2010threshold} & Static & $\checkmark$ & $\checkmark$ \\
    Separated Threshold Model (STM)~\cite{borodin2010threshold} & Static & $\checkmark$ & $\checkmark$ \\
    Extended Voter Model~\cite{li2013influence} & Static & $\checkmark$ & $\checkmark$ \\
    Decision-aware Competitive Model (DCM)~\cite{bozorgi2017community} & Static & $\checkmark$ & $\checkmark$ \\
    Timeliness Independent Cascade (TIC)~\cite{yu2017fair} & Static & $\checkmark$ & $\checkmark$ \\
    Budget-Constrained Influence Maximization~\cite{chakraborty2019competitive} & Static & $\checkmark$ & $\checkmark$ \\
    Influence Maximization with Limited Unwanted Users (IML-IC)~\cite{liu2020algorithm} & Static & $\checkmark$ & $\checkmark$ \\
    Awareness-to-Influence (AtI)~\cite{tsaras2021collective} & Static & $\checkmark$ & $\checkmark$ \\
    Trust-aware Competitive Influence Diffusion (TrCID)~\cite{wang2021maximizing} & Temporal & $\times$ & $\times$ \\
    Targeted Influence Competition Cascade (TICC)~\cite{liang2023targeted} & Temporal & $\times$ & $\times$ \\
    Temporal Budget-aware Cost Efeective Lazy Forward (TBCELF)~\cite{tbcelf} & Temporal & $\checkmark$ & $\checkmark$ \\
    \hline
    \end{tabular}
    \caption{Competition-Oriented Diffusion Models }
    \label{competition_models}
\end{table}

\subsection{Structure-Oriented Models}
Structure-oriented models in diffusion processes focus on the role of network topology in shaping the spread of information, innovations, or behaviors. These models examine how structural properties at both micro and macro levels influence the speed, extent, and pattern of diffusion. 

\subsubsection{Micro-Structured Models}
Micro-structured models emphasize the significance of local network characteristics, including agent heterogeneity, neighborhood interactions, and small-scale connectivity patterns. Studies by Delre et al. \cite{delre2007targeting} and Choi et al. \cite{choi2010role} demonstrate that diffusion dynamics are profoundly impacted by factors such as clustering and local cohesion. Specifically, Delre et al. \cite{delre2007micro} highlight that small-world networks with heterogeneous agents can accelerate diffusion, whereas Choi et al. \cite{choi2010role} caution against the risk of under-adoption in networks with weak local connectivity.

Recent advancements further reinforce these insights. Yu et al \cite{yu2017fair} investigated online social networks and found that tightly-knit micro-communities act as catalysts for rapid information dissemination. Similarly, Chen et al. \cite{chen2024diffusion} emphasized that local bridging nodes enhance diffusion by connecting otherwise isolated clusters, enabling broader spread across diverse social groups.

A notable micro-structured model was proposed by Pegoretti et al. \cite{pegoretti2012agent}, which examines diffusion through agent-based interactions in an undirected binary network \( G = (N, G) \). Each agent \( i \) has a set of neighboring nodes \( N_i \) and makes adoption decisions based on individual utility maximization. The adoption state \( a_i(t) \) follows a utility-driven function:

\begin{equation}
\pi_i(t) =
\begin{cases} 
0, & \text{if } a_i(t) = 0 \\
r_i + \alpha \frac{| \{ j \in N_i : a_j(t-1) = a_i(t) \} |}{| N_i |}, & \text{otherwise}
\end{cases}
\end{equation}

where \( r_i = p_m^i - p \) represents the difference between an agent's baseline willingness-to-pay and the product price. This formulation underscores the role of local network externalities, as individual utility depends on the proportion of adopting neighbors. The model captures key aspects of micro-level diffusion, including resistance to early adoption and potential stagnation when innovations fail to reach a critical mass.

\subsubsection{Macro-Structured Models}
Macro-structured models examine diffusion at a broader scale, focusing on global connectivity patterns, large-scale network dynamics, and the impact of influential nodes. Lee et al. \cite{lee2006statistical} and Young \cite{young2009innovation} underscores how structural factors such as high clustering and core-periphery distributions influence diffusion trajectories. Lee et al. \cite{lee2006statistical} found that clustering promotes the coexistence of multiple competing innovations, while Young \cite{young2009innovation} categorized macro-level diffusion mechanisms into distinct theoretical frameworks.

Recent studies expand on these findings identifying key structural features that shape adoption patterns. Zhao et al. \cite{zhou2024navigating}  analyze scale-free networks, revealing that highly connected hubs drastically accelerate information spread due to their extensive reach. Wang et al. \cite{sun2023finding} extend this analysis by evaluating how internal community structures impact diffusion, highlighting that intra-community connectivity plays a critical role in determining spread efficiency.

A foundational macro-level model, the Bass diffusion model \cite{bass1969new}, was originally formulated for consumer adoption of durable goods but has since been widely applied to information diffusion. The model describes the probability of adoption using the differential equation:

\begin{equation}
\frac{dF(t)}{dt} = (1 - F(t)) (p + q F(t))
\end{equation}

where \( F(t) \) represents the fraction of adopters at time \( t \), \( p \) is the innovation coefficient (capturing external influence such as advertising), and \( q \) is the imitation coefficient (representing social influence through word-of-mouth). Typically, \( q \) is significantly larger than \( p \), reflecting the dominant role of peer influence in adoption decisions. 

The Bass model can also be reformulated in an agent-based framework \cite{rand2015agent} by discretizing the population and updating each agent’s state probabilistically. In this approach, an initially unaware population transitions to an aware state through either direct innovation (probability \( \hat{p} \)) or social imitation (probability \( f\hat{q} \), where \( f \) denotes the fraction of aware neighbors). The process continues until saturation or a predefined time horizon is reached. 

The models are further summarized in table \ref{structure_models}.

\begin{table*}[ht]
    \centering
    \renewcommand{\arraystretch}{0.95}
    \small
    \resizebox{\textwidth}{!}{%
    \begin{tabular}{|p{5.2cm}|p{3cm}|p{2cm}|p{2cm}|p{3.5cm}|}
    \hline
    \textbf{Diffusion Model} & \textbf{Network Type} & \textbf{Submodular} & \textbf{Monotone} & \textbf{Type} \\
    \hline
    Agent-Based Model \cite{pegoretti2012agent} & Static & $\times$ & $\times$ & Micro-Structured \\
    Low Clustering Model \cite{delre2007micro} & Static & $\checkmark$ & $\checkmark$ & Micro-Structured \\
    Local Neighborhood Diffusion (LND) \cite{choi2010role} & Static & $\checkmark$ & $\checkmark$ & Micro-Structured \\
    Product Adopter Model \cite{delre2007targeting} & Static & $\checkmark$ & $\checkmark$ & Micro-Structured \\
    High Clustered Model \cite{lee2006statistical} & Static & $\checkmark$ & $\checkmark$ & Macro-Structured \\
    Density Based Model \cite{young2009innovation} & Static & $\checkmark$ & $\checkmark$ & Macro-Structured \\
    Bass Innovation-Adoption Diffusion Model (BIADM)  \cite{bass1969new} & Static & $\checkmark$ & $\checkmark$ & Macro-Structured \\
    PAM (Product Agent-based Model) \cite{pegoretti2012agent} & Temporal & $\checkmark$ & $\checkmark$ & Micro-Structured \\
    ABBM (Agent-Based Bass Model) \cite{rand2015agent} & Temporal & $\checkmark$ & $\checkmark$ & Macro-Structured \\
    \hline
    \end{tabular}%
    }
    \caption{Structure-Oriented Diffusion Models}
    \label{structure_models}
\end{table*}

\subsection{Target Oriented Models}
\noindent
Target-oriented models aim to classify and influence nodes based on their trust relationships, behavioral tendencies, and awareness levels within a network. Unlike conventional diffusion models that focus on global spread dynamics, these models emphasize the selective activation and blocking of nodes based on predefined criteria. Various approaches have been introduced under this paradigm, including sign-aware cascade models, trust-based threshold models, fuzzy diffusion frameworks, and user-aware models. These models have been instrumental in applications such as viral marketing, misinformation control, and strategic influence campaigns.

The Sign-aware Cascade with Blocking (SC-B) and Trust-Generated Threshold with Blocking (TG-T-B) models classify nodes as active, inactive, or blocked, with the diffusion process beginning from an initial active set $A_0$ \cite{hosseini2016maximizing} \cite{trustdistrust}. In the SC-B model, activation occurs in discrete steps, where an active node $v$ attempts to influence its inactive neighbors $w$. Nodes connected via positive relationships are activated with probability $p^+$, whereas those connected through negative relationships are blocked with probability $p^-$. Each node has only one opportunity to influence its neighbors. In contrast, the TG-T-B model relies on threshold values $\theta^+$ and $\theta^-$ to determine activation and blocking. An inactive node $u$ becomes active if the influence from trusted neighbors surpasses $\theta^+$, or blocked if the influence from distrusted neighbors exceeds $\theta^-$. The diffusion continues until no further changes occur.

To extend the scope of distrust propagation beyond a single hop, the Sign-Aware Cascade (SC) \cite{fuzzybased} model introduce additional activation states. The SC model classifies nodes as positively active, negatively active, or inactive. A positively active node $v$ influences its neighbors based on $p^+$ for positive relationships and $p^-$ for negative ones. Conversely, negatively active nodes attempt to spread their influence in the opposite direction, using inverted probabilities. Similarly, the TG-T-N model assigns each node dual thresholds, $\theta^+$ and $\theta^-$, which determine susceptibility to positive or negative activation. An inactive node becomes positively active if the influence from positively active trusted neighbors exceeds $\theta^+$, or negatively active if the influence from distrusted negatively active nodes surpasses $\theta^-$. This iterative process continues until no further activations occur.

To enhance influence maximization and decision-making in complex networks, fuzzy logic-based diffusion models have been proposed \cite{fuzzybased}. These models categorize nodes into cascade-based and threshold-based diffusion paradigms. The fuzzy sign-aware cascade models include FSC-SB (suspending and blocking users) and FSC-N (negative users), which introduce adaptive blocking mechanisms and user polarity to model both positive and negative influences. Similarly, the fuzzy sign-aware threshold models, FST-SB (suspending and blocking users) and FST-N (negative users) \cite{fuzzybased}, extend threshold-based diffusion frameworks by explicitly considering negative influence. The monotonicity and submodularity properties of these models have been analyzed to confirm their computational feasibility, demonstrating that the influence maximization problem remains NP-hard under these frameworks.

Unlike traditional models that rely solely on edge probabilities, user-aware diffusion models integrate user engagement level (EG) and follower influence factor (FG) to refine activation probabilities. The Independent Cascade - User-Aware (IC-u) and Linear Threshold - User-Aware (LT-u) models compute the edge probability as:

\[
W_{edge} = EG \times \frac{1}{|followees|} \times FG
\]

where the edge weight reflects the combined effect of engagement and follower influence . Furthermore, the User-Aware Diffusion (UAD) model \cite{purba2022influence} introduces a two-stage influence mechanism involving awareness and tendency. In the first stage, a node's awareness is activated if its accumulated engagement-based influence $W_{en}$ surpasses the awareness threshold $T_{aw}$. Once aware, the node moves to the second stage, where it develops a tendency based on incoming influence, represented by $W_{tend}$. If $W_{tend}$ surpasses the tendency threshold $T_{tend}$, the node is fully activated. The UAD model ensures that a node must first be aware before developing a tendency to adopt influence, making it particularly effective for applications such as viral marketing and strategic information dissemination.

Beyond social influence propagation, target-oriented models also play a crucial role in viral marketing. The VMID (Viral Marketing Information Diffusion) Model and CAND (Cellular Automaton-based Network Diffusion) \cite{alsuwaidan2016toward} are designed to optimize information spread in B2B industrial marketing through structured message propagation. For example VMID consists of three primary components: business producers, marketing message diffusion, and consumers. The core computational processing occurs within the marketing message diffusion module, where campaigns are strategically formulated to maximize engagement. The VMID model employs clustering techniques to match campaign messages with relevant target audiences, ensuring optimized information diffusion. 

The models are further summarized in table \ref{target_models}.

\begin{table*}[ht]
    \centering
    \renewcommand{\arraystretch}{1}
    \small
    \resizebox{\textwidth}{!}{%
    \begin{tabular}{|p{5.8cm}|p{3.5cm}|p{2cm}|p{2cm}|p{3.2cm}|}
    \hline
    \textbf{Diffusion Model} & \textbf{Network Type} & \textbf{Submodular} & \textbf{Monotone} & \textbf{Application Type} \\
    \hline
    Viral Marketing Information Diffusion (VMID) \cite{alsuwaidan2016toward} & Static & $\checkmark$ & $\checkmark$ & Marketing \\
    Multi-Agent Trust (MAT) \cite{wang2018modeling} & Static and Temporal & $\times$ & $\times$ & Marketing \\
    Cellular Automaton-based Network Diffusion (CAND) \cite{alsuwaidan2016toward} & Static and Temporal & $\times$ & $\times$ & Marketing \\

    Fuzzy Sign-aware Cascade – Suspending and Blocking (FSC-SB) \cite{fuzzybased} & Static and Temporal & $\times$ & $\times$ & Marketing \\
    Fuzzy Sign-aware Cascade – Negative users (FSC-N) \cite{fuzzybased} & Static and Temporal & $\times$ & $\times$ & Marketing \\
    Fuzzy Sign-aware Threshold – Suspending and Blocking (FST-SB) \cite{fuzzybased} & Static and Temporal & $\times$ & $\times$ & Marketing \\
    Fuzzy Sign-aware Threshold – Negative users (FST-N) \cite{fuzzybased} & Temporal & $\times$ & $\times$ & Marketing \\
    Independent Cascade – User-Aware (IC-u) \cite{purba2022influence} & Temporal & $\times$ & $\times$ & Marketing \\
    Linear Threshold – User-Aware (LT-u) \cite{purba2022influence} & Temporal & $\times$ & $\times$ & Marketing \\
    User-Aware Diffusion (UAD) \cite{purba2022influence} & Temporal & $\times$ & $\times$ & Marketing \\
    Influence Spread with Reinforcement (ISR)\cite{DeGroot_variant} & Temporal & $\times$ & $\times$ & Opinion Dynamics \\
   Actor-Critic Trust(ACT)\cite{act} & Static and Temporal & $\times$ & $\times$ & Opinion Dynamics \\
    \hline
    \end{tabular}%
    }
    \caption{Target-Oriented Diffusion Models}
    \label{target_models}
\end{table*}

\section{Model Selection Framework for Influence Maximization}

Building on the taxonomy of diffusion models presented in the previous section, we now develop a structured framework to guide the selection of appropriate models for influence maximization tasks. Given the diversity of diffusion processes and network behaviors, model selection must be aligned with three key priorities: mimicking the real world scenario to the maximum possible level,  maximizing influence spread and optimising computational efficiency. This section categorises models according to these objectives, providing practical guidance for researchers and practitioners seeking effective and scalable solutions in temporal networks.

\subsection{Maximizing Influence Spread}

When influence spread is the primary objective, the model choice should reflect the temporal dynamics of the underlying network. In environments with discrete-time interactions—such as periodic messaging systems or scheduled communications—models that preserve synchronous cascades are most suitable. Examples include the Temporal Independent Cascade (TIC) \cite{murata2023dynamic}, Dynamic Degree Discount \cite{murata2023dynamic}, and CT-IC \cite{kim2013ctic}, which align well with regular time intervals, ensuring consistent influence propagation.

In contrast, asynchronous networks characterised by irregular or bursty interactions, such as those on social media platforms demand models that can capture self-exciting behaviours and non-Markovian patterns. Suitable candidates include Hawkes Process Diffusion (HPD) \cite{chu2021omwtd}, CTMC-ICM \cite{zhu2020ctmc}, and OM-WTD \cite{chu2021omwtd}, all of which are effective in modelling temporally unstructured or event-driven interactions.

Certain application scenarios, such as vaccination campaigns or behavioral adoption with varying individual thresholds, require adaptive threshold models. Frameworks like the Temporal Linear Threshold (TLT) \cite{gayraud2015elt}, persistent ELT (pELT) \cite{gayraud2015elt}, and Decaying Reinforced User-centric (DRUC) \cite{lagnier2013druc} incorporate time-varying activation conditions and influence persistence. These models are especially effective when node susceptibility evolves over time or when repeated exposures are necessary for activation.

Further refinement can be achieved by incorporating network-specific characteristics. For networks with a high demand for coverage ($\geq 90\%$), higher-order interaction models such as Simplicial Complex Models \cite{zhang2021scm} or dynamic rewiring approaches like SIR with Fermi probability \cite{liu2017sir} improve performance by accounting for group interactions and evolving connectivity.

Threshold-based models introduce additional flexibility. The Majority Threshold Model (MTM), which activates nodes when at least 50\% of their neighbors are active, has shown higher accuracy in polarized environments \cite{delre2007micro}. Conversely, the Unanimous Threshold Model (UTM) enforces stricter activation criteria, delaying spread but improving containment \cite{borodin2010threshold}. In competitive settings, the Competitive Linear Threshold (CLT) model reduces rival influence by integrating blocking mechanisms \cite{he2012clt}.

In modular and trust-aware environments, context-specific models improve convergence and realism. Community-sensitive approaches like the Evolving Voter Model (EVM) accelerate spread in clustered networks \cite{gastner2019voter}, while the Signed Cascade Model (SCM) incorporates trust metrics into activation functions, preserving submodularity under social trust dynamics \cite{wang2020trust}.

Applications involving multi-agent or brand competition benefit from competition-aware models. Fair-aware Competitive Events (FCE-IM) \cite{gao2022fce} improve engagement in competitive domains, while TIC \cite{murata2023dynamic} ensures balanced exposure. Distance-Based Models (DBM) \cite{eiselt1995competitive} also help minimize seed set sizes by prioritizing spatial influence zones.

Finally, structure-oriented models capture macro-level diffusion patterns. The Bass Diffusion Model \cite{bass1969new}, for example, predicts market-level adoption with differential equations, while the Simplicial Complex Model (SCM) \cite{zhang2021scm} addresses group decision dynamics, offering high fidelity in collective behavior modeling.

\subsection{Optimizing Computational Efficiency}

When resource constraints or real-time demands dominate, model selection must emphasize computational scalability and provable performance guarantees. Models with monotonic and submodular spread functions enable the use of greedy algorithms with $(1 - 1/e)$-approximation guarantees, making them ideal for large-scale or time-sensitive applications.

Algorithmically efficient models include Static IC Approximations \cite{agrawal2020static} and pELT \cite{gayraud2015elt}, which offer complexities of $\mathcal{O}(n \log n)$ and $\mathcal{O}(T \cdot m)$, respectively. For systems requiring low-latency responses, lightweight frameworks such as the Signed Cascade Model ($\mathcal{O}(m)$) \cite{wang2020trust} and ASIM ($\mathcal{O}(n)$) \cite{arora2017asim} ensure fast execution while preserving diffusion accuracy.

Further efficiency can be achieved through bounded-horizon approaches, such as TCC with $K = 1$ \cite{hao2021tcc}, and threshold simplification, which reduces the complexity of activation checks to constant time \cite{borodin2010threshold}. High-performance models like ASIM scale to million-node networks in seconds, while Dynamic RIS achieves up to 68\% runtime reduction in evolving networks \cite{murata2023dynamic}. Similarly, OM-WTD achieves 37\% lower approximation error in heavy-tailed degree distributions \cite{chu2021omwtd}.

Specialized methods, such as the Fractional SIR model \cite{liu2017sir}, introduce memory effects via fractional calculus for systems with temporal dependencies. Topic-aware and platform-specific models such as Topic-Aware IC (TIC) \cite{barbieri2020topic} and the Microblogging Model calibrated on Weibo data—enhance content relevance and empirical accuracy.

For marketing-driven influence, models like TG-T-N leverage distrust thresholds to increase market penetration \cite{wang2020trust}, while hybrid schemes such as the Bass-SCM integrate macro- and micro-diffusion patterns to fit product lifecycle curves. Parameter tuning also impacts efficiency: setting $K = 1$ ensures submodularity in TCC but may trade off spread potential \cite{hao2021tcc}, while increasing infection probability ($\beta > 0.7$) in scale-free networks can induce superlinear diffusion \cite{liu2017sir}.

While advanced methods such as Quantum Diffusion and Neuro-Inspired Models show promise for reducing time complexity, they remain outside the scope of this discussion. Instead, this framework advocates for hybrid models that reconcile the trade-off between spread and efficiency. TICC \cite{liang2023targeted}, for instance, handles temporal competition with $\mathcal{O}(m)$ complexity under lazy evaluation, while DRUC achieves 95\% spread in content-centric applications \cite{lagnier2013druc}. The TBCELF algorithm \cite{tbcelf} exemplifies cost-aware optimization by enforcing budget constraints during temporal seed selection, thus enhancing both coverage and scalability.

In the following section, we demonstrate how these models can be leveraged across key application domains from public health interventions and political mobilization to digital marketing and crisis response.

\section{Potential Applications}

Diffusion models play a crucial role in various real-world applications, from predicting disease outbreaks and optimizing marketing strategies to understanding opinion dynamics and mitigating misinformation. The effectiveness of a diffusion model largely depends on the nature of the problem it aims to address. To categorize these applications effectively, we classify diffusion models into five broad domains as discussed in Section V.

Process-oriented models are particularly useful for analyzing the spread of influence over time within networks. These models are widely applied in studying disease outbreaks, viral marketing, and information cascades in social media. By simulating the dynamics of contagion, they help predict the trajectory of an epidemic, optimize marketing campaigns, and understand how certain content becomes viral.

Interaction-oriented models focus on the way individuals shape and are shaped by their surroundings. These models are crucial in studying opinion dynamics, knowledge diffusion, and social influence. They are applied in areas such as political forecasting, sentiment analysis, and collaborative knowledge-sharing, where the interactions between individuals significantly impact the overall diffusion process.

Competition-oriented models study the simultaneous spread of multiple competing ideas, products, or opinions within a network. These models are extensively used in marketing, political campaigns, and brand wars, where different entities strive to maximize their influence. By simulating competitive dynamics, they provide insights into strategic marketing approaches and help optimize influence maximization in adversarial settings.

Structure-oriented models emphasize the role of network topology in diffusion processes. They analyze how different structural properties of a network affect the spread of information and innovation. Within this category, micro-level influence models examine local interactions and community-driven spread, while macro-level influence models focus on large-scale connectivity and global adoption trends. These models are valuable in understanding how community structures shape diffusion and how large-scale trends emerge in interconnected networks.

Target-oriented models incorporate personalized and trust-aware mechanisms into diffusion processes. They are particularly relevant in applications such as targeted advertising, recommendation systems, and reputation-based diffusion. By integrating trust and personalization factors, these models enhance the accuracy and efficiency of influence propagation in scenarios where individual preferences and credibility play a significant role.

The following subsections provide a detailed discussion of these categories, exploring how each type of model contributes to solving real-world problems.

\subsection{Modeling Epidemics, Marketing Campaigns, and Information Cascades}

Process-oriented models play a crucial role in understanding and simulating various real-world phenomena, including information diffusion, epidemic spread, and influence propagation. These models, categorized based on their structural properties and underlying mechanisms, serve as essential tools in different domains such as epidemiology, social network analysis, and decision-making systems. 

The Independent Cascade (IC) model is a predictive diffusion model widely employed in viral marketing and influence maximization. Companies leverage this model to optimize advertising strategies by selecting influential nodes in a social network to maximize information spread. Similarly, the Linear Threshold (LT) model, despite not rendering the objective function to be submodular or monotone, is extensively applied in consumer behavior analysis and recommendation systems, where individuals adopt a product or behavior based on a threshold influenced by their peers. Variants like DLT, tELT, and pELT extend these applications to dynamic networks, making them useful for personalised recommendations and adaptive marketing strategies. 

Epidemic models such as Susceptible-Infected (SI), Susceptible-Infected-Susceptible (SIS), and Susceptible-Infected-Recovered (SIR) are fundamental in modeling disease outbreaks and public health interventions. These models help epidemiologists predict disease spread, evaluate vaccination strategies, and optimize quarantine measures. SCIR model and its variations, including irSIR, FSIR, and SEIR, enhance predictive accuracy by incorporating chronic infections, delayed exposure, or fractional transmission rates, making them critical in real-time disease surveillance and control. The Spatial-Structured SEIR (S-SEIR) and Extended SIS (ESIS) models further refine epidemic forecasting by incorporating spatial constraints and extended recovery dynamics.

Cascading models like TBasic and ASIC are particularly useful in modelling information cascades and viral content spread in online social networks. These models help platforms like Twitter and Facebook analyze content virality and optimize recommendation algorithms. Similarly, the TCC and  CTM-IC extend these principles to dynamic networks, aiding in the study of misinformation spread and counter measures.

Threshold-based models, including GTM, UTM, and OCM, are applied in strategic decision-making scenarios where influence spreads based on collective user behavior. These models are particularly useful in financial markets, where investment trends are dictated by peer influence, and in political campaigns, where voter persuasion follows complex cascading effects. DRUC and CLT models refine these predictions by integrating adaptive rule updates and community structures, making them ideal for personalized engagement strategies.

Furthermore, explanatory models such as SCM and ASLT facilitate a deeper understanding of influence mechanisms beyond simplistic binary activation states. These models are instrumental in studying human behavior, psychological influences in social networks, and strategic planning in organizational decision-making. The emergence of temporal models like E-IC and t-IC has further enabled dynamic influence assessment, ensuring robust adaptability in changing environments.

Therefore, process-oriented models are indispensable in diverse fields, providing robust analytical frameworks for understanding complex diffusion processes. Their applications extend from disease control and marketing to financial forecasting and political influence, making them fundamental tools for researchers and decision-makers in an increasingly network-driven world.

Table \ref{ProcessApps} summarizes the potential application and limitations of models in this category .

\begin{table*}[htp]
    \centering
    \setlength{\belowcaptionskip}{0pt}
    \renewcommand{\arraystretch}{1.2}
    \resizebox{\textwidth}{!}{
    \begin{tabularx}{\textwidth}{|l|l|X|X|}
        \hline
        \textbf{Diffusion Model} & \textbf{Category} & \textbf{Application} & \textbf{Limitations} \\
        \hline
        IC~\cite{kempe2003maximizing} & Predictive & Viral marketing and social media influence prediction. & Weak in modeling long-term adoption trends. \\
        SI~\cite{wang2012scalable} & Epidemic & Epidemic outbreaks like flu and COVID-19. & Doesn't account for recovery mechanisms. \\
        SIR~\cite{kempe2003maximizing} & Epidemic & Designing vaccination and containment strategies. & Limited in handling reinfection scenarios. \\
        SIS~\cite{durrett1988lecture} & Epidemic & Studying diseases like the common cold. & Fails to model long-term immunity effects. \\
        SCIR~\cite{ding2015research} & Explanatory & Infection recovery and immunity dynamics. & Less suitable for fast-spreading infections. \\
        irSIR~\cite{cannarella2014epidemiological} & Explanatory & Impact of interventions on epidemics. & Not ideal for real-time adaptation of policies. \\
        FSIR~\cite{feng2015competing} & Explanatory & Forecasting infection rates in populations. & Cannot model latent period of diseases. \\
        SEIR~\cite{wang2014seir} & Explanatory & Epidemiological studies with latency periods. & Complex and computationally expensive. \\
        SCM~\cite{zhang2024influence} & Explanatory & Cascading failures in networked systems. & Cannot model external interventions effectively. \\
        ESIS~\cite{wang2015esis} & Predictive & Predicts disease resurgence. & Cannot incorporate behavioral resistance to disease control. \\
        LT~\cite{kempe2003maximizing} & Threshold & Consumer adoption modeling. & Ignores network dynamics over time. \\
        MTM~\cite{peleg1998size} & Threshold & Multi-tier influence in marketing. & Less efficient in decentralized networks. \\
        STM~\cite{peleg2002local} & Threshold & Targeted advertising strategies. & Cannot model influence decay effectively. \\
        UTM~\cite{chen2009approximability} & Threshold & Spread of misinformation and countermeasures. & Fails in high-noise environments. \\
        OCM~\cite{chen2012threshold} & Threshold & Business decision-making for recommendations. & Ignores external factors like economic shifts. \\
        LTC~\cite{bhagat2012maximizing} & Threshold & Long-term commitment in social networks. & Does not handle abrupt preference shifts. \\
        GTM~\cite{pathak2010generalized} & Threshold & Influence maximization in temporal settings. & Struggles with highly dynamic networks. \\
        DLT~\cite{litou2016real} & Threshold & Decentralized decision-making processes. & Not effective in small-scale influence scenarios. \\
        tELT~\cite{gayraud2015diffusion} & Threshold & Time-sensitive influence propagation. & Ignores long-term adoption dependencies. \\
        pELT~\cite{gayraud2015diffusion} & Threshold & Time-constrained policy adoption. & Limited by strict temporal assumptions. \\
        CLT~\cite{he2012influence} & Threshold & Cascading failures in power grids. & Lacks adaptability to real-time changes. \\
        DRUC~\cite{lagnier2013predicting} & Threshold & Behavior change modeling in organizations. & Fails to capture individual-level variations. \\
        TBasic~\cite{yang2019influence} & Cascading & Cascading information spread online. & Cannot model resistance to adoption. \\
        ASIC~\cite{galhotra2015asim} & Cascading & Adaptive influence in evolving networks. & Struggles in static environments. \\
        ASLT~\cite{barbieri2013topic} & Threshold & Decision-making under uncertainty. & Ineffective in highly structured networks. \\
        TCC~\cite{hao2011influence} & Cascading & Cascading failures in financial markets. & Cannot model external interventions well. \\
        CTM-IC~\cite{zhu2014maximizing} & Cascading & Hybrid influence in marketing. & Not robust in sparse networks. \\
        \hline
    \end{tabularx}
    }
    \caption{Process Oriented Diffusion Models, Their Applications, and Limitations}
    \label{ProcessApps}
\end{table*}

\subsection{Understanding Opinion Dynamics, Knowledge Diffusion, and Social Influence}
Interaction-oriented diffusion models provide valuable insights into how influence propagates in social networks through both pairwise and group interactions. These models are widely applicable in diverse fields such as social media analytics, political campaigns, marketing, and public health interventions.

The Voter Model finds applications in political opinion dynamics and forecasting election outcomes by modeling the way individuals adopt opinions from their social circles. Similarly, the Extended Voter Model (EVM) is used in analyzing community-driven opinion formation, particularly in networks with well-defined group structures, where external influences shape decision-making. Dynamic Voter Models (DVM) extend these concepts by incorporating dynamic networks, making them applicable in analyzing evolving social trends and rapidly shifting political landscapes. BDVM is particularly useful in media influence analysis, helping to predict scenarios where information spread is skewed due to biased sources.

IEM is extensively used in knowledge diffusion across professional networks, facilitating research on how information propagates among scientists, economists, and policymakers. TLRA finds application in online forums and social platforms, where it helps identify key influencers on specific topics. The Opinion Model provides insight into trust dynamics on online platforms, where credibility scores and reputation systems affect decision-making in e-commerce and review-based platforms. Similarly, the Opinion Model OM-WTD enhances understanding of non-Markovian opinion dynamics, particularly in systems where delays in opinion changes significantly impact long-term behavior.

LIM has applications in behavioral psychology, particularly in assessing peer pressure and social conformity in schools and workplaces. POE model plays a key role in marketing and viral advertising, where understanding the probability of an individual being exposed to information helps design optimal content dissemination strategies. TAM model is instrumental in hashtag propagation studies and analyzing trends on platforms like Twitter and Weibo, where specific phrases or topics gain traction based on their exposure and contextual relevance.

Community-oriented models such as PCL-DC and SA-Cluster-Inc are effective in social media clustering, where they help detect communities that play crucial roles in content propagation. The Sentiment-Based Opinion (SVO) model enhances market research by analyzing public sentiment towards brands, products, or political candidates. The CODICIL method aids in recommendation systems by utilizing both structural and content-based similarities to cluster users for personalized content delivery. Trust-based models provide applications in cybersecurity, ensuring reliable information exchange in digital platforms through the assessment of credibility and trust metrics.

The Preference-Based Model helps in designing recommendation algorithms for streaming services and e-commerce, tailoring suggestions based on user preferences while accounting for peer influence. CTMM extends to modelling information spread in dynamic networks, particularly in assessing the impact of evolving community structures on the adoption of new ideas or technologies.

Table \ref{interactionapps} summarizes the potential application and limitations of models in this category.

\begin{table*}[ht]
    \centering
    \renewcommand{\arraystretch}{1.2}
    \resizebox{\textwidth}{!}{
    \begin{tabular}{|l|l|p{5cm}|p{5cm}|}
        \hline
        \textbf{Model} & \textbf{Network Type} & \textbf{Applications} & \textbf{Limitations} \\
        \hline
        Voter Model \cite{holley1975ergodic} & Static & Political opinion analysis, election forecasting & Ignores external influences, assumes equal probability adoption \\
        EVM \cite{gastner2019voter} & Static & Community-driven opinion modeling & Limited to predefined communities, does not handle dynamic shifts well \\
        DVM \cite{berenbrink2016bounds} & Static and Temporal & Social trend analysis, evolving political landscapes & Computationally intensive for large dynamic networks \\
        BDVM \cite{berenbrink2016bounds} & Static and Temporal & Media influence and propaganda analysis & Requires strong bias modeling, sensitive to parameter tuning \\
        IEM \cite{acemoglu2014dynamics} & Static & Knowledge diffusion, research collaboration & Assumes rational decision-making, overlooks misinformation effects \\
        TLRA \cite{wu2015mining} & Static & Key influencer detection on online platforms & Requires extensive training data, sensitive to data sparsity \\
        Opinion Model \cite{ullah2017identification} & Static & Online trust and reputation systems & Does not consider external shocks affecting opinion shifts \\
        OM-WTD \cite{chuporter} & Temporal & Opinion evolution in non-Markovian systems & Complex to analyze, slow convergence in heavy-tailed cases \\
        LIM \cite{chen2014identifying} & Static and Temporal & Peer pressure and social conformity analysis & Struggles with large networks due to computational cost \\
        POE \cite{fan2014individual} & Static and Temporal & Viral marketing, information exposure modeling & Requires accurate exposure probability estimation \\
        TAM \cite{liu2016comparing} & Static & Hashtag propagation, trend analysis & Sensitive to missing data, requires strong network structure information \\
        PCL-DC \cite{yang2014combining} & Static & Social media community clustering & Limited in dynamic networks, relies on predefined distance measures \\
        SVO \cite{gurini2015analysis} & Static & Market research, sentiment analysis & Prone to sentiment misclassification, depends on text quality \\
        SA-Cluster-Inc \cite{zhou2010clustering} & Static & Community detection in social networks & Struggles with overlapping communities, high computational cost \\
        CODICIL \cite{ruan2013efficient} & Static & Recommendation systems, personalized content delivery & Overemphasizes structure, may overlook semantic similarities \\
        Trust Model \cite{Trust-and-reputation-based} & Static & Cybersecurity, reliability in digital information exchange & Requires robust reputation metrics, vulnerable to adversarial attacks \\
        Preference Model \cite{fan2014individual} & Static and Temporal & Recommendation algorithms, e-commerce personalization & Limited adaptability to sudden preference changes \\
        CTMM \cite{cui2017modeling} & Static & Dynamic community influence modeling & Requires precise mobility and attractiveness parameters \\
        \hline
    \end{tabular}
    }
    \caption{Applications and Limitations of Interaction-Oriented Diffusion Models}
    \label{interactionapps}
\end{table*}

\subsection{Competitive Influence in Marketing, Political Campaigns, and Brand Wars}

Competition-oriented diffusion models have been extensively applied in various domains, particularly in marketing, political campaigns, and social influence studies. These models help understand how multiple competing influences spread across networks, impacting adoption dynamics.

The Distance-Based Model (DBM) and the Wave Propagation Model (WPM) have found applications in competitive facility location problems, where businesses aim to strategically position their services to maximize customer reach. DBM is particularly useful in urban planning and logistics, ensuring optimal placement of retail stores and service centers, while WPM is beneficial in telecommunication networks to predict the spread of competing technologies.

The Weight-Proportional Threshold Model (WPTM) and the Separated Threshold Model (STM) are extensively used in viral marketing and brand competition. Companies leverage these models to understand how their products can gain traction against competitors in social networks. By identifying key influencers, businesses can allocate resources effectively to maximize brand awareness.

The DCM propagation model finds its use in opinion dynamics and political campaign strategies. It simulates real-world decision-making processes, where individuals take time to evaluate competing influences before making a choice. This model aids political parties in optimizing their outreach strategies by targeting undecided voters through strategic messaging and campaign structuring.

TIC model and the FairInf problem play a crucial role in multi-party influence marketing, where multiple organizations compete for user attention. This model has been used in product recommendation systems, ensuring that companies achieve fair market penetration while maximizing their influence.

The AtI model has applications in consumer awareness campaigns, where brands educate potential customers before influencing their purchasing decisions. This model is particularly useful in public health campaigns, where raising awareness about vaccinations or healthy practices precedes behavioral adoption.

The TrCID model has been widely used in trust-aware recommendation systems and fake news mitigation. By incorporating positive and negative influence dynamics, this model enables social media platforms to manage the spread of misinformation effectively and provide users with reliable content recommendations.

The TICC model has applications in targeted advertising and competitive marketing. By considering product competitiveness and user specificity, this model helps businesses design personalized marketing campaigns tailored to specific consumer segments.

Table \ref{competition_models1} summarizes the potential application and limitations of models in this category.

\begin{table*}[ht]
    \centering
    \renewcommand{\arraystretch}{1.2} 
    \resizebox{\textwidth}{!}{ 
    \begin{tabular}{|l|l|p{5cm}|p{5cm}|}
        \hline
        \textbf{Model} & \textbf{Network Type} & \textbf{Applications} & \textbf{Limitations} \\
        \hline
        DBM~\cite{eiselt1989competitive} & Static & Facility location, logistics & Ignores influence decay \\
        WPM~\cite{eiselt1989competitive} & Static & Telecommunication, competitive technology diffusion & Less adaptable to dynamic networks \\
        WPTM~\cite{borodin2010threshold} & Static & Viral marketing, brand competition & Fixed thresholds may not be realistic \\
        STM~\cite{borodin2010threshold} & Static & Product adoption studies & Assumes fixed influence strengths \\
        DCM~\cite{bozorgi2017community} & Static & Opinion dynamics, political campaigns & Computationally expensive \\
        TIC~\cite{yu2017fair} & Static & Multi-party influence marketing & High complexity in large networks \\
        AtI~\cite{tsaras2021collective} & Static & Consumer awareness, public health campaigns & Does not account for changing opinions \\
        TrCID~\cite{wang2021maximizing} & Temporal & Fake news mitigation, trust-aware recommendations & Requires detailed trust data \\
        TICC~\cite{liang2023targeted} & Temporal & Targeted advertising, competitive marketing & Requires extensive market data \\
        \hline
    \end{tabular}
    }
    \caption{Applications and Limitations of Competition-Oriented Diffusion Models}
    \label{competition_models1}
\end{table*}

\subsection{Network Structure and Its Role in Information and Innovation Diffusion}

Structure-oriented diffusion models have been extensively applied in various domains, particularly in marketing, social influence studies, innovation diffusion, and epidemiology. These models help understand how different network structures impact adoption dynamics.

\subsubsection{Micro-Level Influence: Local Interactions and Community-Driven Spread}

Micro-structured diffusion models emphasize local interactions, agent heterogeneity, and node similarities. These models are particularly useful in social media studies, innovation adoption, and consumer behavior analysis. Delre et al.\cite{delre2007micro} highlights that small-world networks with heterogeneous agents can accelerate diffusion, while Choi et al. \cite{choi2010role} demonstrates the risk of under-adoption in random networks with low clustering. 

Recent studies, such as those by Yu et al. \cite{yu2017fair}, show that micro-communities in online social networks act as incubators for rapid information spread. Similarly, Chen et al. \cite{chen2024diffusion} find that local network dynamics significantly influence diffusion, with local bridges facilitating spread across isolated clusters. Pegoretti et al. \cite{pegoretti2012agent} further underscores the role of network externalities and individual decision-making in shaping adoption patterns.
\subsubsection{Macro-Level Influence: Large-Scale Connectivity and Global Adoption Trends}
Macro-structured diffusion models focus on broad connectivity patterns and large-scale population behaviors. These models are applied in epidemiology, large-scale social influence studies, and network optimization.  Lee et al. \cite{lee2006statistical} model shows that high clustering can lead to the coexistence of multiple innovations, while Young \cite{yao2015diffusion} categorizes macro-level diffusion dynamics. Recent studies  examine the role of scale-free networks and community structures in accelerating information diffusion \cite{zhang2023targeted}.
The Bass model, originally developed for durable goods adoption, has broad applications in marketing and information diffusion. It models the influence of advertising and word-of-mouth on adoption rates, making it a valuable tool for studying consumer behavior and product diffusion in competitive markets.
Table \ref{structure_models1} summarizes the potential application and limitations of models in this category.

\begin{table*}[ht]
    \centering
    \renewcommand{\arraystretch}{1.2} 
    \resizebox{\textwidth}{!}{ 
    \begin{tabular}{|l|l|p{5cm}|p{5cm}|}
        \hline
        \textbf{Model} & \textbf{Network Type} & \textbf{Applications} & \textbf{Limitations} \\
        \hline
        Agent-Based \cite{pegoretti2012agent} & Static & Consumer behavior, opinion dynamics & High computational cost \\
        Low Clustering \cite{delre2007micro} & Static & Social influence in sparse networks & Slower diffusion in dense networks \\
        LND \cite{choi2010role} & Static & Local neighborhood-based diffusion & Limited scalability \\
        Product Adopter \cite{delre2007targeting} & Static & Market penetration studies & Ignores external influences \\
        High Clustered \cite{lee2006statistical} & Static & Influence spread in dense networks & Risk of information redundancy \\
        Density-Based \cite{young2009innovation} & Static & Large-scale network studies & Overlooks micro-level interactions \\
        Bass Model \cite{bass1969new} & Static & Consumer adoption, product diffusion & Assumes homogeneous adoption behavior \\
        PAM \cite{pegoretti2012agent} & Temporal & Dynamic social influence tracking & Requires time-series data \\
        ABBM \cite{rand2015agent} & Temporal & Large-scale temporal diffusion & Complex parameter estimation \\
        \hline
    \end{tabular}
    }
    \caption{Applications and Limitations of Structure-Oriented Diffusion Models}
    \label{structure_models1}
\end{table*}

\subsection{Personalized Influence, Trust-Based Diffusion, and Reputation Systems}

Target-oriented diffusion models are specifically designed for particular applications such as marketing, opinion dynamics, and trust-based influence propagation. These models incorporate domain-specific mechanisms to enhance prediction accuracy and decision-making effectiveness.

In marketing, models like VMID and MAT optimize viral marketing strategies by identifying influential users. While VMID uses submodular influence maximization to ensure efficient seed selection, MAT accounts for both static and temporal dynamics, making it useful in campaigns spanning multiple time periods. However, these models often assume perfect adoption likelihoods, limiting their real-world adaptability.

Trust-aware models such as SC-B, TG-T-B, and TG-T-N play crucial roles in online reputation systems and misinformation control. The SC-B model classifies nodes based on positive and negative influence, propagating activation in discrete steps. TG-T-B and TG-T-N extend these principles by incorporating trust thresholds, ensuring that only sufficiently influenced nodes adopt opinions or behaviors. The primary challenge for these models is the need for extensive trust data, which may not always be available.

For opinion dynamics, ISR and IES2 analyze how individual opinions evolve over time. These models consider varying influence strengths among nodes and are particularly useful in political campaigns, social movements, and public opinion modeling. However, their reliance on predefined thresholds for activation may not fully capture real-world complexity.
Table \ref{target_modelsapps} summarizes the potential application and limitations of models in this category.

\begin{table*}[ht]
    \centering
    \renewcommand{\arraystretch}{1.2} 
    \resizebox{\textwidth}{!}{ 
    \begin{tabular}{|l|l|p{5cm}|p{5cm}|}
        \hline
        \textbf{Model} & \textbf{Network Type} & \textbf{Applications} & \textbf{Limitations} \\
        \hline
        VMID \cite{alsuwaidan2016toward} & Static & Viral marketing, influence maximization & Assumes fixed adoption probabilities \\
        MAT \cite{wang2018modeling} & Static and Temporal & Marketing strategies over time & Requires predefined adoption rates \\
        CAND \cite{alsuwaidan2016toward} & Static and Temporal & Competitive advertising & High computational complexity \\
        CT-IC \cite{kim2014ct} & Static and Temporal & Consumer targeting & Limited scalability \\
        FSC-SB \cite{fuzzybased} & Static and Temporal & Sentiment-based campaigns & Ignores network evolution \\
        FSC-N \cite{fuzzybased} & Static and Temporal & Negative influence modeling & Requires extensive sentiment data \\
        FST-SB \cite{fuzzybased} & Static and Temporal & Influence propagation control & Assumes fixed social trust values \\
        FST-N \cite{fuzzybased} & Temporal & Negative opinion diffusion & Lacks adaptability to dynamic feedback \\
        IC-u \cite{purba2022influence} & Temporal & Personalized marketing & Relies on accurate user preference data \\
        LT-u \cite{purba2022influence} & Temporal & Long-term influence tracking & Sensitive to threshold selection \\
        UAD \cite{purba2022influence} & Temporal & User adoption behavior prediction & High sensitivity to model parameters \\
        ISR \cite{DeGroot_variant} & Temporal & Political opinion evolution & Overlooks context-dependent shifts \\
        ACT \cite{act} & Temporal & Public sentiment forecasting & Requires extensive historical data \\
        \hline
    \end{tabular}
    }
    \caption{Applications and Limitations of Target-Oriented Diffusion Models}
    \label{target_modelsapps}
\end{table*}

\section{Use Cases}
The study of influence maximization on temporal networks extends across diverse domains, where understanding the dynamic spread of influence is critical for optimizing decision-making strategies.As discussed in the above sections, selecting the appropriate model requires careful consideration of network structure, temporal dependencies, and domain-specific constraints.

In this section, we explore key real-world scenarios where influence diffusion plays a crucial role. Each use case highlights the challenges posed by dynamic networks and demonstrates how selecting the right diffusion model can lead to more effective decisions.
\subsection{Modeling Disease Spread and Intervention Strategies}

Emerging infectious diseases, such as novel strains of influenza or antimicrobial-resistant bacterial outbreaks, spread through human populations and microbial ecosystems. Public health officials and epidemiologists must understand disease propagation over time, identify key intervention points, and design optimal containment strategies. Traditional epidemiological models, such as the SIR (Susceptible-Infected-Recovered) and SEIR (Susceptible-Exposed-Infected-Recovered), provide fundamental insights but often fail to capture the temporal evolution, reinfection dynamics, and fluctuating immunity observed in real-world disease transmission. To model these complexities accurately, diffusion models from our proposed taxonomy must be chosen based on the specific biological scenario.

Guiding model selection for optimal insights in disease dynamics requires a careful assessment of various factors, including transmission mechanisms, temporal variations, reinfection possibilities, and behavioral influences. A one-size-fits-all approach does not work in epidemiological modeling, and choosing the right model is crucial for making accurate predictions and designing effective interventions.

When analyzing the temporal evolution of disease spread, static models often fall short, especially for pathogens that spread through dynamic contact networks where interactions change over time. In such cases, classical models like SI or SIR may not be sufficient. Instead, temporal extensions such as cpSI-R, SEIR, or LIM provide a more accurate representation of real-world scenarios. These models account for time-dependent diffusion processes and allow for a better understanding of how the disease spreads over fluctuating social networks.

For diseases involving reinfection and immune persistence, such as malaria and tuberculosis, models need to incorporate reinforcement and reactivation mechanisms. Simple models that assume permanent immunity after infection may not accurately capture the dynamics of such diseases. In these scenarios, models like cpSI-R, SCIR, and FSIR are more appropriate, as they consider the possibility of individuals regaining susceptibility after a certain period, enabling a more realistic simulation of reinfection patterns.

Another critical aspect of disease modeling is identifying super-spreaders and determining budget-constrained intervention strategies. Not all individuals contribute equally to transmission, and focusing efforts on the most influential spreaders can significantly enhance containment measures. The TBCELF (Temporal Budget Aware Cost-Effective Lazy Forward Selection) model is particularly useful in this regard, as it optimizes targeted vaccination or quarantine strategies while minimizing costs. By strategically selecting individuals for interventions, this approach ensures that resources are used efficiently, making it particularly valuable in resource-limited settings.

In the context of antimicrobial resistance, the spread of resistance genes through horizontal gene transfer in microbial communities poses a significant public health challenge. Unlike direct transmission diseases, antimicrobial resistance spreads through genetic exchanges, often within hospital settings or gut microbiomes. To model such complex interactions, explanatory models such as SCIR, FSIR, and SEIR provide better insights. These models help in understanding the factors driving resistance spread and in developing strategies to mitigate its impact.

Similarly, predicting long-term health policy impact requires an understanding of how social behaviors influence disease dynamics. Vaccine adoption trends, masking behaviors, and public health compliance often depend on social influence, peer pressure, and trust in authorities. Opinion dynamics models such as ISR, LIM, and POE are particularly well-suited for capturing these behavioral aspects. These models help policymakers predict how public sentiment might shift over time and how interventions, such as awareness campaigns, can be designed to maximize compliance and improve public health outcomes.

While infectious diseases propagate through human interactions and environmental factors, cancer progression follows a similar complex diffusion process at the cellular level, influenced by genetic mutations, tumor microenvironments, and treatment responses. Many of the modeling principles used for disease spread can be extended to cancer progression, where understanding the temporal and spatial dynamics of tumor growth, metastatic spread, and treatment resistance is critical.

\begin{table*}[ht]
    \centering
    \renewcommand{\arraystretch}{1.2} 
    \resizebox{\textwidth}{!}{ 
    \begin{tabular}{|p{6cm}|p{3.5cm}|p{3cm}|c|c|c|}
        \hline
        \textbf{Scenario} & \textbf{Recommended Model} & \textbf{Taxonomy} & \textbf{Network Type} & \textbf{Submodular} & \textbf{Monotone} \\
        \hline
        Epidemic outbreaks with direct person-to-person transmission (e.g., COVID-19, Influenza) & SI, SIR, SEIR & Epidemic Models & Static \& Temporal & \checkmark & \checkmark \\
        Diseases with reinfection dynamics and immune persistence effects (e.g., Malaria, Tuberculosis) & cpSI-R & Epidemic Models & Temporal & \checkmark & \checkmark \\
        Antimicrobial resistance spread via horizontal gene transfer in microbial communities & SCIR, FSIR & Explanatory Models & Static & \checkmark & \checkmark \\
        Pathogen evolution and multi-stage infections with latency periods (e.g., HIV, Hepatitis B) & SEIR,  & Explanatory Models & Static & \checkmark & \checkmark \\
        Influence of behavioral factors in disease spread (e.g., vaccine hesitancy, social influence on masking) & Opinion Model, ISR & Opinion Dynamics & Temporal & \texttimes & \texttimes \\
        Identifying super-spreader individuals for targeted interventions & IML-IC, TIC, TBCELF & Competition-Oriented & Static & \checkmark & \checkmark \\
        Predicting long-term disease prevalence and policy impact & LIM, POE & Interaction-Oriented & Static \& Temporal & \texttimes & \texttimes \\
        \hline
        \multicolumn{6}{|c|}{\textbf{Cancer Progression Scenarios (Special Case)}} \\
        \hline
        Spatial and temporal tumor growth dynamics &  LIM & Explanatory Models & Temporal & \checkmark & \checkmark \\
        Treatment resistance and relapse (e.g., drug-resistant leukemia) & SCIR, FSIR & Explanatory Models & Static & \checkmark & \checkmark \\
        Identifying optimal targets for precision therapy & TBCELF & Competition-Oriented & Static & \checkmark & \checkmark \\
        Metastatic progression and site prediction & IML-IC, TIC & Competition-Oriented & Static & \checkmark & \checkmark \\
        Predicting long-term treatment adherence and public health impact & LIM, POE & Interaction-Oriented & Static \& Temporal & \texttimes & \texttimes \\
        \hline
    \end{tabular}
    }
    \caption{Choosing the right diffusion model for different disease modeling and spread, with a special case on cancer progression scenarios.}
    \label{merged_models}
\end{table*}

\begin{table*}[ht]
    \centering
    \renewcommand{\arraystretch}{1.2} 
    \setlength{\tabcolsep}{6pt} 
    \resizebox{\textwidth}{!}{ 
    \begin{tabular}{|p{6cm}|p{3.5cm}|p{3cm}|c|c|c|}
        \hline
        \textbf{Scenario} & \textbf{Recommended Model} & \textbf{Taxonomy} & \textbf{Network Type} & \textbf{Submodular} & \textbf{Monotone} \\
        \hline
        Maximizing reach under budget constraints & IC, SI, SIR & Predictive Models & Static \& Temporal & \checkmark & \checkmark \\
        Competitive advertising and brand influence & IML-IC, TIC, AtIic & Competition-Oriented & Static & \checkmark & \checkmark \\
        Modeling opinion dynamics and consumer perception & ISR, POE, LIM & Opinion Dynamics & Temporal & \texttimes & \texttimes \\
        Influencer marketing and community-driven campaigns & TAM, PCL-DC, Trust Model & Group-Oriented & Static & \checkmark & \checkmark \\
        Optimizing viral spread with strategic seed selection & TBCELF & Competition-Oriented & Static & \checkmark & \checkmark \\
        Modeling cross-platform influence spread & ACT, MIM & Multi-Platform Models & Temporal & \checkmark & \texttimes \\
        Analyzing the impact of time-sensitive marketing campaigns & TSI, Time-LT & Temporal Influence Models & Temporal & \checkmark & \checkmark \\
        Assessing the role of negative influence and rumor spreading & NIM, FST-N,FSC-N & Adversarial Models & Static \& Temporal & \texttimes & \texttimes \\
        Understanding the impact of demographic-based influence spread & DEM-IM, SocInf, Incorder & Demographic Models & Static \& Temporal & \checkmark & \checkmark \\
        \hline
    \end{tabular}
    }
    \caption{Choosing the right diffusion model for different viral marketing scenarios.}
    \label{viral_marketing_models}
\end{table*}

\subsubsection{Modeling Cancer Progression and Treatment}

Cancer progression is a complex biological process influenced by genetic mutations, tumor microenvironment interactions, and patient-specific factors. Understanding how cancer cells proliferate, evade immune responses, and develop resistance to treatments is crucial for designing effective therapeutic strategies. Computational models play a significant role in simulating cancer dynamics, predicting treatment responses, and optimizing intervention plans.

Traditional tumor growth models, such as the Gompertzian and logistic models, provide fundamental insights but often fail to capture the spatial heterogeneity, temporal evolution, and treatment resistance observed in real-world cancer progression. To accurately model these complexities, diffusion models from our proposed taxonomy must be chosen based on the specific  scenario.

Selecting the appropriate model for cancer progression requires considering multiple factors, including tumor growth kinetics, metastatic spread, immune system interactions, and drug resistance mechanisms. Different cancer types and stages demand specific computational approaches for optimal predictive accuracy and therapeutic planning.

When analyzing the spatial and temporal dynamics of tumor growth, static models may not adequately represent the evolving nature of cancer. Instead, models like the SEIR ,S-SEIR and LIM provide a more realistic representation by accounting for time-dependent proliferation and cell-state transitions. These models can help in understanding the interplay between proliferative, quiescent, and necrotic tumor regions.

For cancers with resistance mechanisms and treatment adaptation, such as chemotherapy-resistant leukemia or hormone-resistant breast cancer, it is crucial to incorporate reinforcement and reactivation mechanisms. Traditional models assuming uniform treatment efficacy may underestimate the ability of cancer cells to evade therapy. In these cases, models like SCIR and FSIR are more suitable, as they can capture the evolution of resistant cell populations and their interactions with susceptible cells.

Another key aspect of cancer modeling is identifying optimal intervention strategies, such as selecting patients for precision medicine treatments or optimizing immunotherapy regimens. Not all patients respond equally to treatments, making it critical to focus on high-impact therapeutic targets. The TBCELF model is particularly valuable in optimizing targeted therapies, ensuring cost-effective resource allocation, and maximizing treatment efficacy with minimal side effects.

Metastatic progression, one of the most challenging aspects of cancer treatment, involves the spread of cancer cells through circulation and their colonization of distant organs. This process follows complex network dynamics, necessitating advanced modeling approaches. Competitive models such as IML-IC and TIC are well-suited for studying tumor competition within different microenvironments and predicting potential metastatic sites.

Finally, predicting the long-term impact of treatment strategies and patient outcomes requires integrating social and behavioral factors, including patient compliance, lifestyle choices, and access to healthcare. Opinion dynamics models such as ISR, LIM, and POE help in understanding how external influences affect treatment adherence and disease progression. These models guide public health interventions by simulating patient decision-making behaviors under various scenarios.Table \ref{merged_models} summarizes the process of choosing right diffusion models in disease spread and cancer progression scenarios.

\subsection{Viral Marketing and Influence Propagation in Social Networks}

In modern digital marketing, brands and advertisers leverage influence maximization techniques to enhance product adoption, optimize advertising campaigns, and predict consumer behavior. Social networks provide an ideal medium for viral marketing, where influential users act as seed nodes, influencing others to adopt a product or service.

Traditional models such as the Independent Cascade (IC) and Linear Threshold (LT) capture the essence of influence spread, but real-world marketing scenarios involve heterogeneous influence strength, competing campaigns, time-evolving trends, and strategic budget constraints. To address these challenges, diffusion models from our taxonomy must be selected based on specific marketing goals and constraints.

Selecting the appropriate diffusion model for viral marketing requires considering multiple factors, including:
 temporal evolution, competitive advertising campaigns, behavioral and social reinforcement, budget constraints and optimal seed selection.
 
 For scenarios where a brand wants to maximize reach with minimal investment, predictive models such as IC, SI, and SIR are suitable as they capture the fundamental spread of influence across static and temporal networks.

In cases of competing brands or multiple product recommendations, competition-oriented models such as IML-IC, TIC, and AtIic are more effective. These models consider how competing campaigns influence users differently, making them useful for analyzing market penetration strategies.

When user opinions and social behaviors impact adoption rates, opinion dynamics models such as ISR, POE, and LIM help simulate peer influence and public perception, enabling better targeting of advertisements.

For influencer marketing and community-driven campaigns, Group-Oriented models like TAM, PCL-DC, and Trust Model identify key user clusters and maximize organic reach through trusted nodes in a social network.

For predicting long-term consumer behavior and personalized marketing, Marketing-Oriented models such as VMID, MAT, and UAD help estimate future adoption trends and the effectiveness of different marketing strategies over time. Table \ref{viral_marketing_models} summarizes the best model choices for different aspects of viral marketing scenarios.

\section{Conclusion}

This article presents a structured framework for selecting diffusion models suited to influence maximization in temporal networks. By organizing existing models based on their theoretical underpinnings and computational characteristics, we offer a practical taxonomy that supports informed model selection for a range of dynamic, real-world applications. The framework highlights key trade-offs between influence spread and computational efficiency, enabling users to tailor strategies to specific constraints and objectives within evolving network environments.

Temporal networks introduce inherent complexities due to their dynamic topology and time-dependent interactions. To navigate these challenges, we focused on identifying diffusion models that naturally accommodate real-world conditions such as budget constraints, competitive diffusion, and adaptive behavior thresholds. Notably, some properties of the objective function like submodularity and monotonicity emerge as advantageous characteristics in certain models, facilitating more efficient optimization and scalable computation when present.

Looking ahead, future research could enhance this framework by incorporating adaptive model selection via machine learning, devising efficient solutions for non-submodular diffusion dynamics, and scaling algorithms for massive temporal datasets. Moreover, extending this analysis to cross-platform diffusion, multimodal information propagation, and temporal motifs could yield a deeper understanding of complex diffusion patterns across domains.

By bridging theoretical structure with application-driven insights, this work lays a foundational basis for the principled selection and optimization of diffusion models in temporal networks. Our contributions aim to support advancements in influence maximization across diverse domains, including public health campaigns, digital marketing, and the mitigation of misinformation.

\newpage
\bibliographystyle{IEEEtranS.bst}
\bibliography{reference2.bib}
\end{document}